\documentclass[a4paper]{amsart}

\usepackage{lineno}
\usepackage[hidelinks]{hyperref}

\usepackage{amsmath}
\usepackage{amssymb}
\usepackage{amsthm}
\usepackage{hyperref}
\usepackage{subfig}
\usepackage{graphicx}
\usepackage{graphbox} 
\usepackage[left=3cm, right=3cm, bottom=3cm, top=3cm]{geometry}

\usepackage{float}
\usepackage{xcolor}
\hypersetup{
    colorlinks,
    linkcolor={red!50!black},
    citecolor={blue!50!black},
    urlcolor={blue!80!black}
}
\usepackage{amsaddr}

\theoremstyle{plain} 

\DeclareMathOperator{\tr}{tr}

\DeclareMathOperator{\Div}{Div}

\DeclareMathOperator{\sym}{sym}

\DeclareMathOperator{\cof}{cof}

\newcommand{\vect}[1]{\mathbf{#1}}
\newcommand{\tens}[1]{\mathsf{#1}}
\newcommand{\onablap}{\overline{\nabla}_\parallel}
\newcommand{\onablao}{\overline{\nabla}_\perp}
\newcommand{\onabla}{\overline{\nabla}}

\newcommand{\bb}{\mathcal{B}}

\usepackage{mathtools}
\usepackage{pgfplots}
\pgfplotsset{/pgf/number format/use comma,compat=newest}

\renewcommand\epsilon{\varepsilon}

\newtheorem*{remark*}{\bf Remark}
\usepackage{upgreek}

\begin{document}
\title{The F\"{o}ppl--von K\'{a}rm\'{a}n equations of elastic plates with initial stress}
\author{P. Ciarletta, G. Pozzi, D. Riccobelli}
\email{pasquale.ciarletta@polimi.it}
\address{MOX -- Dipartimento di Matematica, Politecnico di Milano, piazza Leonardo da Vinci 32, 20133 Milano, Italy.}

\begin{abstract}
Initially stressed plates are widely used in modern fabrication techniques, such as  additive manufacturing  and UV lithography, for their tunable morphology by application of external stimuli.  In this work, we propose a formal asymptotic derivation of the F\"{o}ppl--von K\'{a}rm\'{a}n equations for an elastic plate with initial stresses,  using the constitutive theory of nonlinear elastic solids with initial stresses under the assumptions of incompressibility and material isotropy.  Compared to existing works, our approach allows to determine the morphological transitions  of the elastic plate without prescribing the underlying target metric of the unstressed state of the elastic body.
We explicitly solve the derived FvK equations in some physical problems of engineering interest, discussing how the  initial stress distribution  drives the emergence of spontaneous curvatures within the deformed plate.  The proposed mathematical framework can be used to tailor shape on demand, with applications in several engineering fields ranging from soft robotics  to 4D printing.
\end{abstract}

\maketitle

\section{Introduction}

The F\"{o}ppl--von K\'{a}rm\'{a}n (FvK) equations are a set of nonlinear partial differential equations describing the large deflection of linear elastic plates \cite{foppl1921vorlesungen, karman1907festigkeitsprobleme}. They can be derived as a formal asymptotic expansion of the three-dimensional filed theory of linear elasticity in the limit of large displacements and small
strains, and associated with specific boundary conditions \cite{goodier1938problems,friedrichs1961boundary,gol1962derivation,ciarlet1980justification,ciarlet2006equations}.

The FvK equations are notably difficult to solve, but they proved to be very useful to give theoretical insights in many physical problems where in-plane and out-of-plane unknowns can be decoupled \cite{ben1997crumpled,cerda2003geometry,audoly2008buckling,audoly2010elasticity,vella2017indentation}. The research interest in FvK equations has been recently reinvigorated by the technological possibility to fabricate shape-morphing devices using soft active materials \cite{fernandes2010multiparameter,felton2014method,zhang2014defects,gladman2016biomimetic,bowick2017non,van2018mechanics,tao2021morphing}.  These morphable plates have  applications in several engineering fields, ranging from soft robotics \cite{feinberg2007muscular,hines2017soft} to the design of biomimetic structures \cite{gladman2016biomimetic}. In particular, morphological transitions can be realized by controlling the geometric frustration of a soft plate by swelling \cite{kim2010dynamic,pezzulla2015morphing,van2017growth}, surface accretion \cite{zurlo2017printing}, optothermal stimuli in nematic elastomers \cite{Mihai_2020}, and surface tension in nano-plates \cite{yue2017modified}. For these purposes, FvK equations have been derived in cases where geometrical incompatibilities arise and the undeformed configuration of the elastic plates is no longer free of initial  stresses \cite{dervaux2009morphogenesis}. 

The formal asymptotic expansion leading to the FvK model has been rigorously derived as \mbox{$\Gamma$-limit} of the three dimensional elastic problem \cite{Friesecke_2006, neukamm2013derivation}. This analysis has been recently extended to the case of a pre-strained plate \cite{lewicka2011foppl,Bhattacharya_2016,Ricciotti_2017,Bola_os_2020}. The existing approaches account for the geometrical frustration using additive or multiplicative decomposition of the deformation gradient for describing the spatial distribution of residual strains given by the underlying non-Euclidean metric \cite{klein2007shaping,dervaux2008morphogenesis,efrati2009elastic,efrati2013metric}.
However, the FvK equations with pre-strains require the prescription of the incompatible metric of the virtual relaxed state, whilst in many practical cases the distribution of residual strains remains unknown. Since a stress-free configuration cannot be physically attained by non-invasive techniques, it is more suitable to consider a more general theoretical framework where the elastic energy explicitly depends on the spatial distribution of the initial stress in the reference configurations, without the need to prescribe a stress-free state \cite{ciarletta2016residual}.   

In the following, we present a formal asymptotic derivation of the FvK equations for an elastic plate with initial stresses. In Section 2, we introduce the constitutive theory of nonlinear elastic solids with initial stresses under the assumptions of incompressibility and material isotropy. In Section 3, we introduce the scaling assumption for the geometrical parameters and the initial stress components and we derive the FvK equations for an initially stressed elastic plate. In Section 4, we explicitly solve the derived FvK equations in some physical problems, discussing how the initial stress concentration may drive the emergence of spontaneous curvatures within the deformed plate. Concluding remarks are finally summarized in Section 5.

\section{Constitutive theory for initially stressed elastic materials}

Let us consider a  body that occupies a simply connected domain  $\bb_{\boldsymbol \tau}$  in its reference configuration. We use Cartesian unit vectors $({\bf E}_X,{\bf E}_Y,{\bf E}_Z)$ and $({\bf e}_x,{\bf e}_y,{\bf e}_z)$ in the reference and spatial configurations, respectively.   Let ${\bf X} =  X \vect{E}_X + Y \vect{E}_Y + Z \vect{E}_Z$ being the material position vector.
In this undeformed configuration, the body has an initial  stress, meaning that its Cauchy stress tensor is not vanishing. We denote the initial stress tensor by $\boldsymbol \tau$, where ${\boldsymbol \tau}:\bb_{\boldsymbol{\tau}}\rightarrow \mathcal{S}(\mathbb{R}^3)$, and $\mathcal{S}(\mathbb{R}^n)$ is the set of the self-adjoint $\tens{L}:\mathbb{R}^n\rightarrow\mathbb{R}^n$, where $\tens{L}\in\mathcal{L}(\mathbb{R}^n)$ is a linear application.  
In order to enforce the material balance of linear and angular momentum, this initial stress is such that :
\begin{equation}
\Div {\boldsymbol \tau} = {\bf 0}, \qquad {\boldsymbol \tau}= {\boldsymbol \tau}^T,  \qquad {\rm in} \quad \bb_{\boldsymbol \tau},
\label{eqres}
\end{equation}
where $\Div$ is the material divergence operator.
We remark that if the body is residually stressed, i.e.  the zero-traction boundary condition  ${\boldsymbol \tau} {\bf N} = {\bf 0}$ applies to the whole boundary  $\partial \bb_{\boldsymbol \tau}$ with material unit normal ${\bf N}$,   ${\boldsymbol \tau}$ must be inhomogeneous and have zero average over the volume in $\bb_{\boldsymbol \tau}$ in force of the mean value theorem \cite{hoger85}.

Let ${\bf x}= \boldsymbol{\varphi}({\bf X})=x\vect{e}_x+y\vect{e}_y+z\vect{e}_z$ be the spatial material vector, so that  $\boldsymbol \varphi: \bb_{\boldsymbol \tau} \rightarrow\bb$ be the one-to-one mapping to the deformed configuration $\bb$ and $\tens{F}=\partial \boldsymbol\varphi/\partial{\bf X}$ be the deformation gradient.  In the following, we deal with incompressible materials, imposing the constraint  $J= \det \tens{ F}= 1$.
We further assume that the body possesses a perfectly elastic response, defining a strain energy density $\Psi$ per unit of reference volume as \cite{shams,merodio13}:
\begin{equation}
\Psi= \Psi(\tens{F}, {\boldsymbol \tau}).
\end{equation}
By standard arguments, the second Piola-Kirchhoff $\tens{S}$ and Cauchy ${\boldsymbol \sigma}$ stress tensors reads:
\begin{equation}
\tens{S}= \frac{\partial \Psi}{\partial \tens{F}}(\tens{F},\,
{\boldsymbol \tau})\tens{F}^{-T}-p\tens{F}^{-1}\tens{F}^{-T}, 
\qquad {\boldsymbol \sigma}=\tens{F} \tens{S}\tens{F}^T, \label{stresses}
\end{equation}
where $p$ is the Lagrange multiplier enforcing the incompressibility constraint. In particular,   when $\tens{F}$ is equal to the identity tensor $\tens{I}$, the Cauchy stress must be equal to the initial stress tensor $\boldsymbol \tau$, so that: 
\begin{equation}
{\boldsymbol \tau}= \frac{\partial \Psi}{\partial \tens{F}}(\tens{I},\,{\boldsymbol \tau})-p_{\boldsymbol \tau} \tens{I}, \label{res-stress}
\end{equation}
where $p_{\boldsymbol \tau}$ is the value of $p$ in $\bb_{\boldsymbol \tau}$.
If we further assume that the elastic response is invariant after the application of a rigid-body motion, the strain energy can be expressed as a function of the  three invariants
of the right Cauchy-Green tensor $\tens{C}=\tens{F}^T \tens{F}$, the three invariants of the initial stress tensor ${\boldsymbol \tau}$, plus their four mixed invariants \cite{spencer1971part,merodio13}. Since we are interested in developing  a FvK theory, we are interested in considering the minimal constitutive response that takes into account for geometric nonlinearities.
Accordingly, we consider in the following the constitutive response of a pre-stressed Neo-Hookean material, that is given by \cite{gower15}
\begin{equation}
  \Psi(\tens{F},\,\tau) = \frac{1}{2} \left(\tr ({\boldsymbol \tau}\tens{C})+ r \tr\tens{C} -E \right),
  \label{psinh}
\end{equation}
where $E$ is the Young modulus of the unstressed material and $r$ is the real root of:
\begin{equation} 
 r^3+I_{\boldsymbol\tau 1} r^2+I_{\boldsymbol\tau 2} r+ I_{\boldsymbol\tau 3}-\left(\frac{E}{3}\right)^3=0,
 \label{root}
\end{equation}
with $ I_{\boldsymbol \tau 1}= \tr {\boldsymbol \tau}$, $I_{\boldsymbol \tau 2}= \tfrac{1}{2}[(I_{\boldsymbol \tau 1}^2 -\tr({\boldsymbol \tau}^2)]$, $  I_{\boldsymbol \tau 3}= \det {\boldsymbol \tau}$.
The Eq.~\eqref{root} is obtained by inverting the constitutive relation \eqref{stresses} and imposing the incompressibility constraint after a repeated application of the Cayley-Hamilton theorem.
Enforcing the compatibility condition \eqref{res-stress} for the initial stress, we also find that $p_{\boldsymbol \tau}=r$.
Such an energy guarantees that the material properties are not affected by the initial stress distribution, for details see \cite{Riccobelli_2019}.

From \eqref{stresses}, the constitutive equation for the initial stressed Neo-Hookean material with strain energy given by \eqref{psinh} is:
\begin{equation}
\tens{S}= r\tens{I}-p \tens{C}^{-1} + {\boldsymbol \tau},
\qquad {\boldsymbol \sigma}=r\tens{B}-p \tens{I} + \tens{F}{\boldsymbol \tau}\tens{F}^T, \label{stressesnh}
\end{equation}
where $\tens{B}=\tens{F}\tens{F}^T$ is the left Cauchy-Green tensor.

Neglecting the presence of volume bulk forces, the equilibrium conditions in the reference configuration read:
\begin{equation}
\Div \tens{S} = \vect{0}, \qquad \tens{S}= \tens{S}^T,  \qquad \text{in} \quad \bb_{\boldsymbol \tau},
\end{equation}
while the zero-traction conditions at the boundary gives $\tens{S}\vect{N}=0$ at $\partial \bb_{\boldsymbol \tau}$. In the following, we will use these constitutive assumptions to derive  a FvK theory of elastic plates with initial  stress.

\section{The F\"{o}ppl--von K\'{a}rm\'{a}n equations for initially stressed plates}
In this section we first make the geometric assumptions of the FvK theory,  using dimensional analysis to justify the asymptotic development. We later derive the constitutive equations using a variational argument, assuming different scaling for the initial stress components.

\subsection{Geometric assumptions and asymptotic analysis }

\begin{figure}[!t]
\centering
\includegraphics[width=\textwidth]{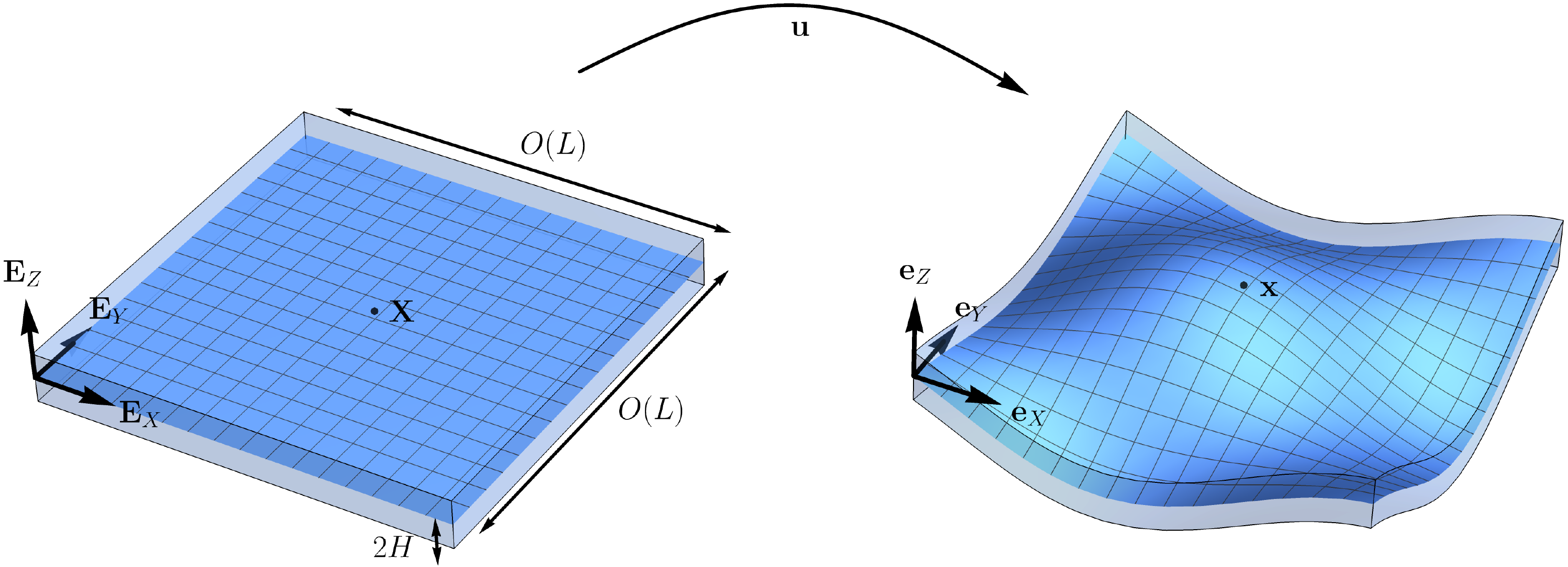}
\caption{Sketch of the reference (left) and current (right) configuration of the elastic plate with initial stress. The blue surfaces represent the mid-section of the plate.}
\label{fig1}
\end{figure}

We consider an elastic plate with reference configuration  $\bb_{\boldsymbol \tau}=S^m \ \times \ [-H,H]$, where $S^m$ is a closed subset in $ \mathbb{R}^2$, as sketched in Figure~\ref{fig1}.
In particular we require that there exist a point $\vect{P}\in S^m$ such that $B_L(\vect{P})\subseteq S^m$ and $L$ is much larger than the thickness $2H$ of the plate, so that $\epsilon=H/L\ll 1$, where $B_L(\vect{P}) = \{\vect{Y} \in \mathbb{R}^2\;|\;\|\vect{Y}-\vect{P}\|< L\}$. The dimensionless parameter $\epsilon$ will be used to perform an asymptotic expansion.
In this thin geometry, we introduce the following dimensionless variables:
\begin{equation}
\overline{X}=X/L, \quad \overline{Y}=Y/L, \quad \overline{Z}=Z/H, \qquad \overline{\vect{X}} = \overline{X}\vect{E}_X + \overline{Y}\vect{E}_Y + \overline{Z}\vect{E}_Z.\label{adim}
\end{equation}
Similarly, let $\overline{\nabla}$ be the gradient operator with respect to the dimensionless variables.

Let ${\bf u}= {\bf x}-{\bf X}$ be the displacement vector describing the deformation of the plate.
We denote by $\vect{u}_\parallel$ the projection of $\vect{u}$ on the plane containing $S^m$, while $w$ indicates the orthogonal component of $\vect{u}$. More explicitly
\begin{equation}
\vect{u} = \vect{u}_\parallel + w \vect{E}_z,\qquad
w = \vect{u}\cdot\vect{E}_Z = H \overline{w},\qquad \vect{u}_\parallel = \mathbb{P}\vect{u} = \vect{u} - w\vect{E}_Z = H \overline{\vect{u}}_\parallel,
\end{equation}
where $\mathbb{P} = \tens{I} - \vect{E}_Z\otimes\vect{E}_Z$ is the projection operator, while $\overline{\vect{u}}_\parallel$ and $\overline{w}$ are the dimensionless counterpart of $\vect{u}_\parallel$ and $w$, respectively. From now on, for a generic quantity $f$ we denote by $\overline{f}$ its dimensionless counterpart.

 We assume the classical Kirchhoff hypothesis for the in-plane displacement $\vect{u}_\parallel$:
\begin{equation}
\label{eq:u_parallel}
\begin{aligned}
\vect{u}_\parallel &= H\overline{\vect{u}}_\parallel  = \vect{U}^m - Z \nabla_\parallel w,\\
\overline{\vect{u}}_\parallel &= \overline{\vect{U}}^m - \epsilon \overline{Z} \overline{\nabla}_\parallel \overline{w},
\end{aligned}
\end{equation}
where $\vect{U}^m(\vect{X}_\parallel)= H \overline{\vect{U}}^m (\overline{\vect{X}}_\parallel)=\vect{u}_\parallel (\mathbb{P}\vect{X})  $ represents the in-plane displacement of the point $\vect{X}_\parallel$ of the middle surface, obtained through the projection of $\vect{X}$ on $S^m$, while $\nabla_\parallel$ is the gradient operator with respect to $\vect{X}_\parallel$, namely in Cartesian components $\nabla_\parallel = \left(\partial/\partial X,\,\partial/\partial Y, 0\right)$, while $\nabla_\perp = (0,0,\partial/\partial Z)$.
From the non-dimensionalization proposed in Eq.~\eqref{adim}, we get
\begin{equation}
\label{eq:nabla_nondim}
\nabla_\parallel=\frac{1}{L}\overline{\nabla}_\parallel,\qquad \nabla_\perp=\frac{1}{H}\overline{\nabla}_\perp.
\end{equation}

The out-of-plane component of the displacement vector reads: 
\begin{equation}
\begin{aligned}
w&= H \overline{w}=\xi^m+\epsilon^2 W,\\
\overline{w}&=\overline{\xi}^m+\epsilon^2 \overline{W},
\end{aligned}\label{W}
\end{equation}
where $\xi^m(\mathbf{X}_\parallel)  = H \overline{\xi}^m(\mathbf{X}_\parallel)=w(\mathbb{P}\vect{X})$ is the out-of-plane displacement of the point $\vect{X}_\parallel$ of the middle surface, obtained through the projection of $\vect{X}$ on $S^m$ and $W:\bb_\tau\rightarrow\mathbb{R}^3$. The field $\xi$ governs the out-of-plane deformation up to an order $O(\epsilon^2)$. 

The assumption made in Eqs.~\eqref{eq:u_parallel} and \eqref{W} imposes that the displacement is a pure bending deformation in the $Z$-direction, so that sections that are initially perpendicular to the middle surface undergo a rotation that is driven by the local curvature of the deformed middle surface. For an initially stressed plates, the validity of this scaling requires a further constraint on the scaling of the initial stress tensor, that will be discussed later.

We further assume that the middle surface components can be expanded in powers of $\epsilon$ as follows:
\begin{equation}
\vect{U}^m=\sum_{n=1}^{\infty} \epsilon^n \vect{U}^m_n=H\sum_{n=1}^{\infty} \epsilon^n \overline{\vect{U}}^m_n, \quad \xi^m=\sum_{n=0}^{\infty} \epsilon^n \xi^m_n=H\sum_{n=0}^{\infty} \epsilon^n \overline{\xi}^m_n.
\end{equation}

Using Eq.~\eqref{eq:nabla_nondim}, the deformation gradient can be written as
\begin{equation}
\tens{F} = \tens{I} + \nabla \vect{u} = \tens{I} + \nabla_\parallel \vect{u} + \nabla_\perp \vect{u} = \tens{I} + \epsilon \overline{\nabla}_\parallel\overline{\vect{u}}+ \overline{\nabla}_\perp\overline{\vect{u}},
\end{equation}
since $\overline{\vect{u}} = \overline{\vect{u}}_\parallel + \overline{w} \vect{E}_Z$, substituting Eqs. \eqref{eq:u_parallel} and \eqref{W}, we get
\begin{equation}
\overline{\vect{u}}= \overline{\vect{U}}^m + \overline{\xi}^m \vect{E}_Z - \epsilon \overline{Z} \onablap \overline{\xi}^m+\epsilon^2 \overline{W}\vect{E}_Z+o(\epsilon^2).
\end{equation}
We can compute $\onablap \overline{\vect{u}}$ and $\onablao \overline{\vect{u}}$, obtaining
\begin{equation}
\begin{aligned}
\onablap \overline{\vect{u}} &= \onablap \overline{\vect{U}}^m + \vect{E}_Z\otimes\onablap\overline{\xi}^m-\epsilon \overline{Z}\onablap\onablap\overline{\xi}^m + o(\epsilon),\\
\onablao \overline{\vect{u}} &= - \epsilon \onablap \overline{\xi}^m \otimes \vect{E}_Z + \epsilon^2 \overline{W}_{,\overline{Z}}\vect{E}_Z\otimes\vect{E}_Z + o(\epsilon^2),
\end{aligned}
\end{equation}
where comma denotes the partial derivative. Thus, the right Cauchy-Green strain tensor becomes
\begin{equation}
\label{C}
\begin{aligned}
\tens{C} &= \tens{F}^T\tens{F}=\\
&= (\tens{I} + \epsilon \onablap \overline{\vect{u}} + \onablao \overline{\vect{u}})^T(\tens{I} + \epsilon \onablap \overline{\vect{u}} + \onablao \overline{\vect{u}})=\\
&=\tens{I} + \epsilon \onablap \overline{\vect{u}} + \onablao \overline{\vect{u}} + \epsilon(\onablap\overline{\vect{u}})^T + \epsilon^2(\onablap\overline{\vect{u}})^T \onablap \overline{\vect{u}} + \epsilon(\onablap\overline{\vect{u}})^T\onablao \overline{\vect{u}}+\\
&\qquad\qquad+ (\onablao\overline{\vect{u}})^T + \epsilon(\onablao\overline{\vect{u}})^T \onablap \overline{\vect{u}} + (\onablao\overline{\vect{u}})^T\onablao \overline{\vect{u}}=\\
&=\tens{I}+\epsilon^2\tens{A} + o(\epsilon)^2,
\end{aligned}
\end{equation}
where the tensor $\tens{A}$ is defined as
\begin{equation}
\tens{A} =2\sym\onabla\overline{\vect{U}}-2 \overline{Z}\onabla\onabla\overline{\xi}+(\|\onabla\overline\xi\|^2+2\overline{W}_{,\overline{Z}})\vect{E}_Z\otimes\vect{E}_Z+\onabla\overline{\xi}\otimes\onabla\overline{\xi},
\end{equation}
here we have set $\overline{\vect{U}}  = \overline{\vect{U}}^m_1$, $\overline{\xi}= \overline{\xi}^m_0$ for the sake of simplicity.
At the leading order, the incompressibility constraint reads $\tr\tens{A}=0$, that imposes: 
\begin{equation}
\overline{W}_{,\overline{Z}}=\overline{Z}\overline{\Updelta} \xi- \onabla\cdot\overline{\vect{U}}-\|\overline\nabla \xi\|^2,
\end{equation}
where $\overline{\Updelta} = \onabla\cdot\onabla$ is the Laplace operator.
The tensor $\tens{A}$ affinely depends on $\overline{Z}$. More explicitly, we write
\begin{equation}
\label{eq:A}
\tens{A} = \tens{C}_0 + \overline{Z} \tens{C}_1,
\end{equation}
with
\begin{equation}
\label{eq:C0C1}
\begin{aligned}
\tens{C}_0 &= 2\sym\onabla\overline{\vect{U}}-(\|\onabla\overline\xi\|^2+2\onabla\cdot\overline{\vect{U}})\vect{E}_Z\otimes\vect{E}_Z+\onabla\overline{\xi}\otimes\onabla\overline{\xi},\\
\tens{C}_1 &=2\overline{\Updelta}\overline\xi\vect{E}_Z\otimes\vect{E}_Z-2\onabla\onabla\overline{\xi}.
\end{aligned}
\end{equation}
In Appendix~\ref{appendix}, we report the explicit expression of the tensors in Eq.~\eqref{eq:C0C1} using the canonical basis $(\vect{E}_X,\,\vect{E}_Y,\,\vect{E}_Z)$. 
Finally, for future convenience, we also remind that the asymptotic expansion of the inverse of $\tens{C}$ reads:
\begin{equation}
\tens{C}^{-1}=\tens{I} - \epsilon^2\tens{A} +o(\epsilon^2). 
\label{invC}
\end{equation}

\subsection{Scaling assumptions on the initial stress} 
Under the constitutive assumption given by \eqref{W}, we define the dimensionless initial stress tensor as $\overline{\boldsymbol \tau}= \boldsymbol \tau/E$. We assume the following scaling for the initial stress tensor:
\begin{equation}
\overline{\boldsymbol \tau}=\epsilon^2\overline{\boldsymbol{\tau}}_2  + \epsilon^3 \overline{\boldsymbol{\tau}}_3+\epsilon^4\overline{\boldsymbol{\tau}}_4,
\label{atau}
\end{equation}
with
\begin{equation}
\begin{aligned}
\overline{\boldsymbol{\tau}}_2&=\overline{\tau}_{XX}\vect{E}_X\otimes\vect{E}_X+\overline{\tau}_{XY}(\vect{E}_X\otimes\vect{E}_Y+\vect{E}_Y\otimes\vect{E}_X)+\overline{\tau}_{YY}\vect{E}_Y\otimes\vect{E}_Y,\\
\overline{\boldsymbol{\tau}}_3&= 2\sym\left((\overline{\tau}_{XZ} \vect{E}_X + \overline{\tau}_{YZ} \vect{E}_Y)\otimes\vect{E}_Z\right),\\
\overline{\boldsymbol{\tau}}_4&=\overline{\tau}_{ZZ} \vect{E}_Z\otimes\vect{E}_Z,
\end{aligned}
\end{equation}
where $\overline{\tau}_{ij}=\overline{\tau}_{ij}(\vect{X})=O(1)$, with $i,\,j=X,\,Y,\,Z$. 
Substituting \eqref{atau} into \eqref{eqres}, the equilibrium equation in the undeformed configuration read:
\begin{equation}
\left\{
\begin{array}{l}
\overline{\tau}_{XX,\overline{X}} + \overline{\tau}_{XY,\overline{Y}} + \overline{\tau}_{XZ,\overline{Z}}=0, \\
\overline{\tau}_{XY,\overline{X}} + \overline{\tau}_{YY,\overline{Y}} + \overline{\tau}_{YZ,\overline{Z}}=0,\\
\overline{\tau}_{XZ,\overline{X}} + \overline{\tau}_{YZ,\overline{Y}} + \overline{\tau}_{ZZ,\overline{Z}}=0.  \\
\end{array}
\right.
\label{atau2}
\end{equation} 

If the top and bottom surfaces are free of traction, i.e. $\overline{\tau}_{iZ}(\overline{X},\overline{Y},\pm 1)=0$, the previous equation can be further simplified by integration in the Z-direction:
\begin{equation}
\label{atau3}
\left\{
\begin{array}{l}
\int_{-1}^{1}(\overline{\tau}_{XX,\overline{X}} + \overline{\tau}_{XY,\overline{Y}}) d\overline{Z}=0, \\
\int_{-1}^{1}(\overline{\tau}_{XY,\overline{X}} + \overline{\tau}_{YY,\overline{Y}}) d\overline{Z}=0, \\
\int_{-1}^{1}(\overline{\tau}_{XZ,\overline{X}} + \overline{\tau}_{YZ,\overline{Y}}) d\overline{Z}=0.
\end{array}
\right.
\end{equation}

We define the averaged planar initial stress tensor $\overline{\boldsymbol{\tau}}_m$ as
\begin{equation}
\boldsymbol{\overline{\tau}}_m = \frac{1}{2}\int_{-1}^1 \left[\overline{\tau}_{XX}\vect{E}_X\otimes\vect{E}_X+\overline{\tau}_{XY}(\vect{E}_X\otimes\vect{E}_Y+\vect{E}_Y\otimes\vect{E}_X)+\overline{\tau}_{YY}\vect{E}_Y\otimes\vect{E}_Y\right]\,d\overline{Z}.
\end{equation}
The leading order  of Eq. \eqref{atau3} can be automatically fulfilled by introducing the dimensionless initial Airy stress function $\overline{\chi}_0 :\overline{S}^m\rightarrow\mathbb{R}$, so that
\begin{equation}
\overline{\boldsymbol{\tau}}_m = \cof(\onabla_m\onabla_m\overline{\chi}_0),
\label{achi0}
\end{equation}
where $\onabla_m$ is the dimensionless gradient operator in $\mathbb{R}^2$, $\cof\tens{A}$ is the cofactor the of tensor $\tens{A}$,
For the sake of clarity, using Cartesian coordinates we get
\begin{equation}
\overline{\boldsymbol{\tau}}_m = 
\begin{bmatrix}
\overline{\chi}_{0, \overline{Y}\overline{Y}} &-\overline{\chi}_{0, \overline{X}\overline{Y}}\\
-\overline{\chi}_{0, \overline{X}\overline{Y}} &\overline{\chi}_{0, \overline{X}\overline{X}}
\end{bmatrix}.
\end{equation} 

Under these assumptions, we get  $ I_{\boldsymbol \tau 1}/E= O(\epsilon^2)$, $I_{\boldsymbol \tau 2}/E^2= O(\epsilon^4)$, $  I_{\boldsymbol \tau 3}/E^3= O(\epsilon^8)$. From \eqref{root}, the leading order expression of  the dimensionless term $\overline{r}=r/E$ is:
\begin{equation} 
\overline{r}=\frac{1}{3}\left(1-\epsilon^2\tr\overline{\boldsymbol{\tau}}_2\right) +o(\epsilon^2).
\end{equation}
Using Eqs.~\eqref{stressesnh}, \eqref{invC} and \eqref{atau}, the components of the dimensionless second Piola-Kirchhoff stress tensor $\overline{\tens{S}}= \tens{S}/E$ scale as follows
\begin{equation} 
\vect{E}_\alpha\cdot\overline{\tens{S}}\vect{E}_\beta=O(\epsilon^2), \quad  \vect{E}_\alpha\cdot\overline{\tens{S}}\vect{E}_Z=O(\epsilon^3), \quad  \vect{E}_Z\cdot\overline{\tens{S}}\vect{E}_Z=O(\epsilon^4),\qquad \alpha,\,\beta \in \{X,\,Y\},
 \label{barS}
\end{equation}
where the latter scaling is enforced by imposing that the dimensionless Lagrange multiplier $\overline{p}= p/E$ reads:
\begin{equation} 
\overline{p}= \overline{r} + \frac{\epsilon^2}{3}\vect{E}_Z\cdot\tens{A}\vect{E}_Z +o(\epsilon^2).
\end{equation}
In particular, we remark that \eqref{barS} corresponds to the classical  {FvK} scaling for the stress tensor, and justifies the mathematical soundness of the Kirchhoff assumption for the displacement field in Eqs.~\eqref{eq:u_parallel} and \eqref{W} in the presence of initial  stress  components which scale as in \eqref{atau}. The leading-order components of the stress tensor reads
\begin{equation} 
\overline{\tens{S}}=\left(\overline{\tens{S}}_0 + \overline{Z} \overline{\tens{S}}_1 +\boldsymbol{\tau}_2\right)\epsilon^2+ o(\epsilon^2),
 \label{barS2d}
\end{equation}
with
\begin{equation}
\overline{\tens{S}}_j= \frac{1}{3}\left(\tens{C}_j-\vect{E}_Z\cdot\tens{C}_j\vect{E}_Z\tens{I}\right),\qquad j=0,\,1.
 \label{S0}
\end{equation}
We remark that both $\overline{\tens{S}}_0$ and $\overline{\tens{S}}_1$ are independent of $\overline{Z}$, since they depend on $\overline{\vect{X}}_\parallel$ through $\overline{\xi}$ and $\overline{\vect{U}}$.

\subsection{Variational formulation} 
Let $\overline{\mathcal{E}}=\mathcal{E}/(E H L^2 )$ be the dimensionless counterpart of the elastic energy. From now on, we will refer only to dimensionless variables (unless differently specified) and we drop the use of the superposed line to indicate the non-dimensional quantities.
We now perform  a variational derivation of the generalised {FvK} theory for an initially stressed material by imposing the stationary conditions for the total elastic energy functional $\mathcal{E}$. Neglecting  traction loads at the boundary and body forces for the sake of simplicity, the first variation of the dimensionless energy   reads: 
\begin{equation}
\begin{aligned}
\delta \mathcal{E}=\int_{S^m} \int_{Z=-1}^{1} \frac{1}{2}\tens{S}:\delta \tens{C}\,dZ dS,
\end{aligned}
\label{deltaE}
\end{equation}
where   $dS= dXdY$ and $\tens{A}:\tens{B} = \tr(\tens{A}^T\tens{B})$. From Eqs.~\eqref{C}, \eqref{eq:A}, and \eqref{eq:C0C1}, the increment of the right Cauchy-Green tensor $\tens{C}$ reads:
\begin{equation}
\delta \tens{C}= \left( \sym\nabla\delta\vect{U}-2Z\nabla\nabla\delta\xi+\nabla \delta\xi\otimes \nabla\xi+\nabla\xi\otimes \nabla \delta\xi\right)\epsilon^2+\delta C_{ZZ}\vect{E}_Z\otimes\vect{E}_Z+o(\epsilon^2),
\label{deltaC} 
\end{equation}
where  $\delta C_{ZZ}$ is the increment of $\vect{E}_Z\cdot\tens{C}\vect{E}_Z$. 

Since $\tens{S}_0$ and $\tens{S}_1$ are symmetric and $\tens{S}_0\vect{E}_Z=\tens{S}_1\vect{E}_Z = \vect{0}$, we can introduce the tensors $\tens{S}_{m0}$ and $\tens{S}_{m1}$ which are their projection onto $\mathcal{S}(\mathbb{R}^2)$.
Substituting  Eqs.~\eqref{barS2d} and \eqref{deltaC} into Eq.~\eqref{deltaE}, we get
\begin{equation}
\delta \mathcal{E}= \delta \mathcal{E}_\text{s}+\delta \mathcal{E}_\text{b},
\end{equation}
where $\delta \mathcal{E}_\text{s}$ and $\delta \mathcal{E}_\text{b}$  are the increments due to average planar stretching and bending of the plate, respectively. They are defined as
\begin{gather}
\label{Est1}
\delta \mathcal{E}_{\text{s}} =2\epsilon^4 \int_{S^m} \left(\tens{S}_{m0}+ \boldsymbol{\tau}_m\right):\nabla_m\delta\vect{U}\,dS +o(\epsilon^4),\\
 \label{Ebend1}
 \delta \mathcal{E}_\text{b} =  2\epsilon^4\int_{S^m}\left[ -\left(\frac{1}{3} \tens{S}_{m1}+\tens{M}\right):  \left(\nabla_m\nabla_m\delta\xi\right)+ (\tens{S}_{m0}+\boldsymbol{\tau}_m)\nabla_m\xi\cdot\nabla_m\delta\xi\right]  dS+o(\epsilon^4),
\end{gather}
 where $\tens{M}$ is the dimensionless tensor representing the average bending torques imposed by the initial stress, defined as 
\begin{equation}
\tens{M} =\frac{1}{2}\int_{-1}^1 Z\left[\tau_{XX}\vect{E}_X\otimes\vect{E}_X+\tau_{XY}(\vect{E}_X\otimes\vect{E}_Y+\vect{E}_Y\otimes\vect{E}_X)+\tau_{YY}\vect{E}_Y\otimes\vect{E}_Y\right]\,dZ,
\end{equation}
 which vanishes if $\tau_{\alpha\beta}$ are  even functions of $Z$, with $\alpha,\,\beta\in\{X,\,Y\}$.
 
In the following, we detail such contributions, deriving the corresponding {FvK} equations as the necessary conditions for their extremal values.

\subsubsection{F\"{o}ppl--von K\'{a}rm\'{a}n equation for the average  planar stretch} 
Integrating by parts \eqref{Est1} and neglecting the remainder, we get
\begin{equation}
 \delta \mathcal{E}_\text{s} =-2 \epsilon^4 \int_{S^m} \left[\nabla_m\cdot\left(\tens{S}_{m0}+ \boldsymbol{\tau}_m\right)\right]\cdot\delta\vect{U}\,dS +{\rm b.t.},
\label{Est2}
\end{equation}
where we write $+ \text{ b.t.}$ to indicate boundary terms. Accordingly, imposing $\delta \mathcal{E}_\text{s}=0$ for each admissible variation $\delta \vect{U}$, we get $\nabla_m\cdot\left(\tens{S}_{m0}+ \boldsymbol{\tau}_m\right)=0$. This equation is automatically fulfilled by introducing the Airy stress function $\chi$ as:\footnote{The existence of the Airy stress function can be shown by using some standard theorems of the theory of distributions (see \cite[p.~59, Théorème VI]{schwartz1957theorie}). For multiply connected domains, one could proceed following the procedure proposed in \cite[pp.~61-66]{ciarlet2006equations}.}
\begin{equation}
\tens{S}_{m0}+\boldsymbol{\tau}_m = \cof \nabla_m\nabla_m\chi,
\label{achi}
\end{equation}
Using Eqs.~\eqref{achi0}-\eqref{S0} in Eq.~\eqref{achi}, we finally obtain:
\begin{equation}
\label{fvk1}
\boxed{
\Updelta_m^2 (\chi-\chi_0) + [\xi,\,\xi]=\Updelta_m^2  \chi+ \left([\xi,\,\xi]-C_G\right)= 0,}
\end{equation}
where $\Updelta_m = \nabla_m\cdot\nabla_m$ is the Laplace operator in $\mathbb{R}^2$ and the bracket operator is the Monge-Ampère bilinear form
\begin{equation}
[a,\,b]= \frac{1}{2}\left(\cof\nabla_m\nabla_m a\right) : \nabla_m\nabla_m b,
\end{equation}
and $C_G= \Updelta_m^2 \chi_0$ plays the role of a spontaneous Gaussian curvature imposed by the planar initial stress.

\subsubsection{F\"{o}ppl--von K\'{a}rm\'{a}n equation for the average bending stretch} 

 Using Eq.~\eqref{S0} in Eq.~\eqref{Ebend1} and neglecting the remainder, we integrate by parts, obtaining:
 \begin{equation}
 \delta \mathcal{E}_\text{b} = 2\epsilon^4\int_{S^m}\left[ \frac{4}{9}\Updelta^2_m \xi -\nabla_m\cdot\left(\nabla_m\cdot \tens{M}\right)- \nabla_m\cdot((\tens{S}_{m0}+\boldsymbol{\tau}_m)\nabla_m\xi) \right] \delta \xi  \  dS \ { \rm + \ b.t. \ }
\label{Ebend2}
\end{equation}
By substituting Eq.~\eqref{achi} in  Eq.~\eqref{Ebend2} and imposing $\delta \mathcal{E}_\text{b}=0$ for each admissible variation $\delta \xi$, we finally get:
\begin{equation}
\label{fvk2}
\boxed{
\frac{2}{9}\Updelta^2_m \xi -\frac{1}{2}\nabla_m \cdot \left(\nabla_m \cdot \tens{M}\right) - [\chi,\xi]=\frac{2}{9}\left(\Updelta^2_m \xi - \Updelta_m C_M\right) - [\chi,\xi]=0,}
\end{equation}
where $\Updelta_m C_M= \frac{9}{4}\nabla_m \cdot\left( \nabla_m \cdot \tens{M}\right)$, so that $C_M$ represents the spontaneous mean curvature imposed by the initial stress distribution. In Appendix~\ref{appendix}, we show the explicit expressions of the FvK equations in Cartesian and polar coordinates.

\section{Physical examples}
In this Section we provide few physical examples showing the morphology of the solutions of the  FvK equations \eqref{fvk1}-\eqref{fvk2} for different distributions of the initial stress within the elastic plate. 

\subsection{Planar initial stress}
Let us start by considering a planar distribution of the initial stress. Since ${\boldsymbol \tau}$ does not depend on $Z$, the corresponding average bending torques vanish, i.e. $\tens{M}=\tens{0}$. Accordingly, such an initial stress distribution may impose only a non-zero spontaneous Gaussian curvature $C_G$ in Eq.~\eqref{fvk1}, while $C_M=0$ in Eq.~\eqref{fvk2}.
For the sake of  clarity, we discuss in the following some physical examples exhibiting different spontaneous Gaussian curvatures.

\subsubsection{Positive spontaneous Gaussian curvature}
Let $R,\,\Theta$ be the polar coordinates of the generic point $\vect{X}_\parallel\in S^m$.
Let us first consider an initial stress distribution having an Airy stress function given by
\begin{equation}
 \chi_0(R,\,\Theta)= \frac{c^2}{64}R^4,
\label{chi0pos}
\end{equation}
with $c$ being a characteristic dimensionless stress parameter, and the planar initial stress tensor $\boldsymbol{\tau}_m$ reads:
 \begin{equation}
 \boldsymbol{\tau}_m= \frac{c^2 R^2}{16}\left(\vect{E}_R\otimes\vect{E}_R+3\vect{E}_\Theta\otimes\vect{E}_\Theta\right).
\label{tau0pos}
\end{equation}
\begin{figure}[!t]
\centering
\includegraphics[width=0.5\textwidth]{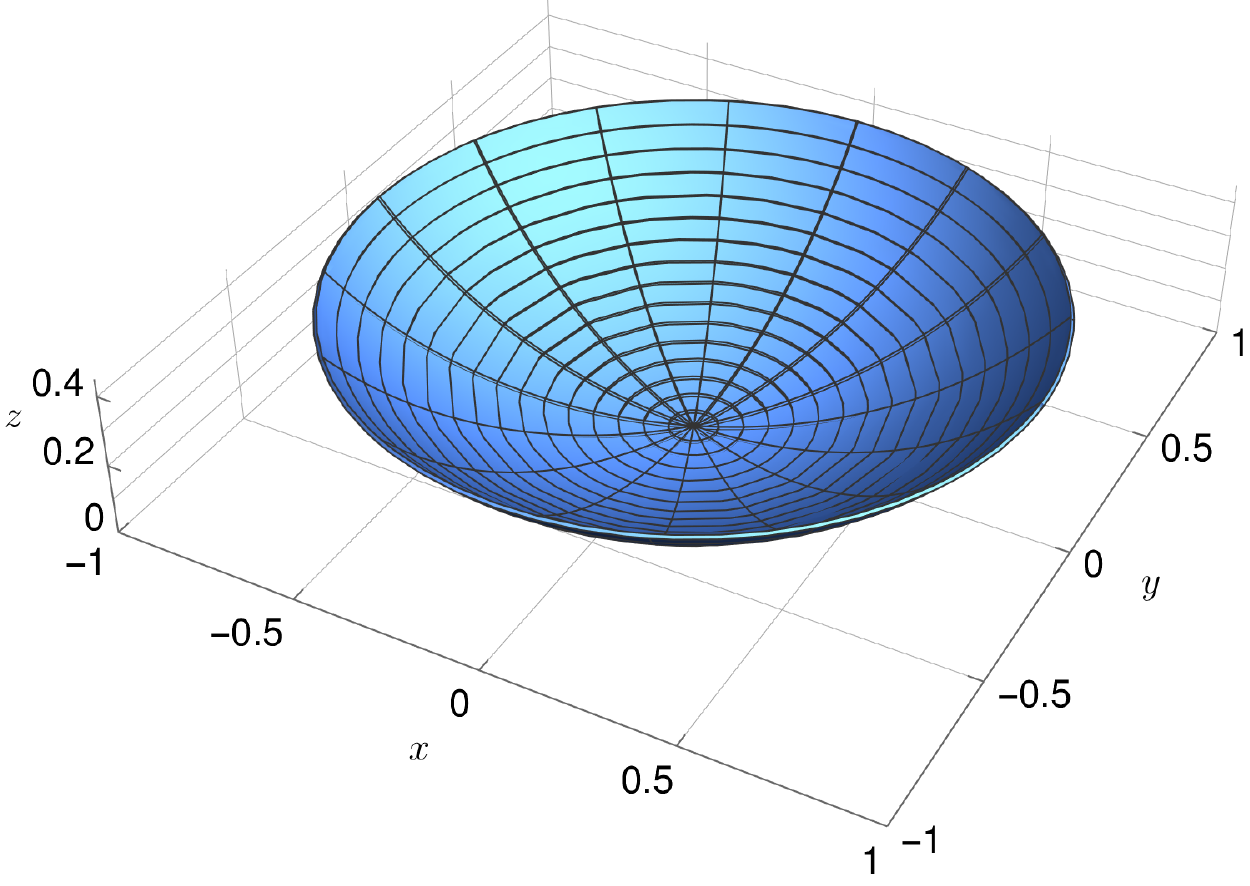}\includegraphics[width=0.5\textwidth]{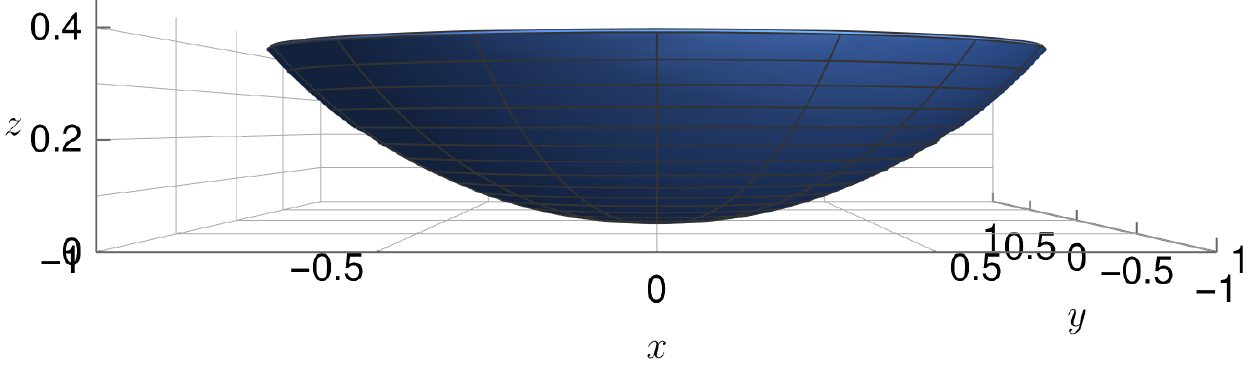}
\caption{Plot of actual buckled configuration of the elastic plate (using dimensional variables) as given by Eq.~\eqref{solposcg} for the initial stress distribution in Eq.~\eqref{tau0pos}. We have set $H=0.01$, $L=1$, and $c=80$, where $S^m = \{\vect{Y}\in\mathbb{R}^2\;|\;\|\vect{Y}\|<1\}$.}
\label{figcgpos}
\end{figure}
From Eq.~\eqref{fvk1},  the spontaneous Gaussian curvature is positive and constant, being:
 \begin{equation}
 C_G= c^2 >0.
\label{cgpos}
\end{equation}
We look for a solution such that the second Piola-Kirchhoff  stress tensor vanishes. In such a case, Eqs.~\eqref{fvk1}-\eqref{fvk2} simplify as
 \begin{equation}
 [\xi,\xi]-c^2= 0, \qquad   \Updelta_m^2 \xi=0,
 \label{fvkposcg}
\end{equation}
where the former is a Monge-Ampére equation. Using polar coordinates, we get
\begin{equation}
\begin{aligned}
[\xi,\xi]&=-\frac{\xi _{,\Theta}^2}{R^4}+\frac{2 \xi _{,R\Theta} \xi _{,\Theta}}{R^3}+\frac{\xi _{,\Theta\Theta} \xi _{,RR}}{R^2}-\frac{\xi _{,R\Theta}^2}{R^2}+\frac{\xi _{,R} \xi _{,RR}}{R},\\
\Updelta_m^2\xi&=\frac{R^2 \xi _{,RRRR}+2 R \xi _{,RRR}-\xi _{,RR}+2 \xi _{,RR\Theta\Theta}}{R^2}+\frac{\xi _{,R}-2 \xi _{,R\Theta\Theta}}{R^3}+\frac{4 \xi _{,\Theta\Theta}+\xi _{,\Theta\Theta\Theta\Theta}}{R^4}.
\end{aligned}
\end{equation}
A solution to Eq.~\eqref{fvkposcg} such that $\xi$ depends only on $R$ is given by
\begin{equation}
 \xi(R)= \frac{c}{2} R^2,
 \label{solposcg}
\end{equation}
which corresponds to a buckling of the plate, as illustrated in Figure~\ref{figcgpos}.

\subsubsection{Negative spontaneous Gaussian curvature}
We now consider the following Airy stress function
 \begin{equation}
 \chi_0(R,\,\Theta)= -\frac{c^2 R^{2 n+4}}{\left(4 n^2+12 n+8\right)^2},\qquad n\in\mathbb{N},
\label{chi0neg}
\end{equation}
with $c$ being a characteristic dimensionless stress parameter.  Accordingly, the planar initial stress reads:
 \begin{equation}
 \boldsymbol{\tau}_m = -\frac{c^2 (2 n+4) R^{2 n+2}}{\left(4 n^2+12 n+8\right)^2}\left(\vect{E}_R\otimes\vect{E}_R + (3 + 2 n) \vect{E}_\Theta\otimes\vect{E}_\Theta\right).
\label{tau0neg}
\end{equation}
\begin{figure}[!t]
\centering
\includegraphics[width=0.5\textwidth]{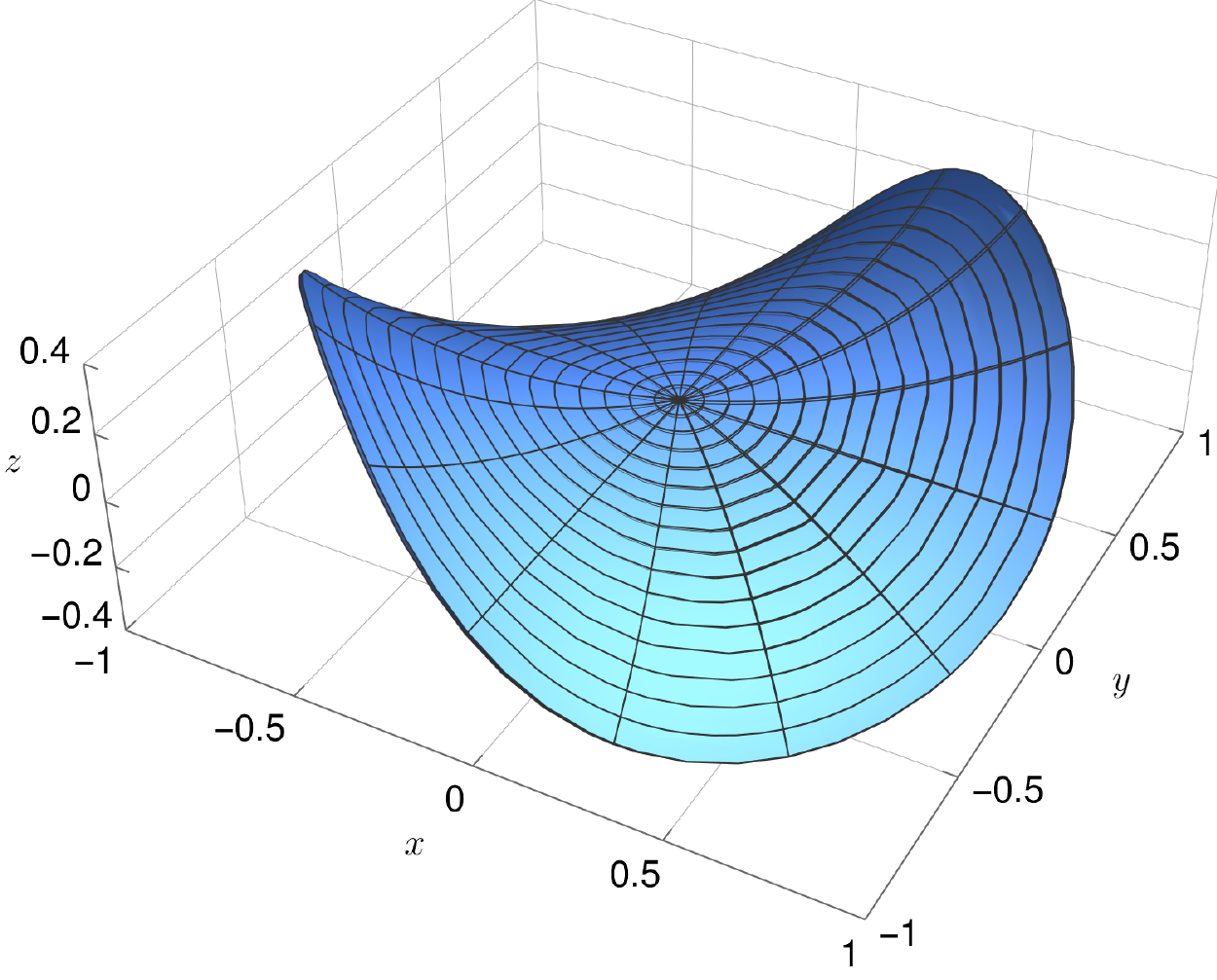}\includegraphics[width=0.5\textwidth]{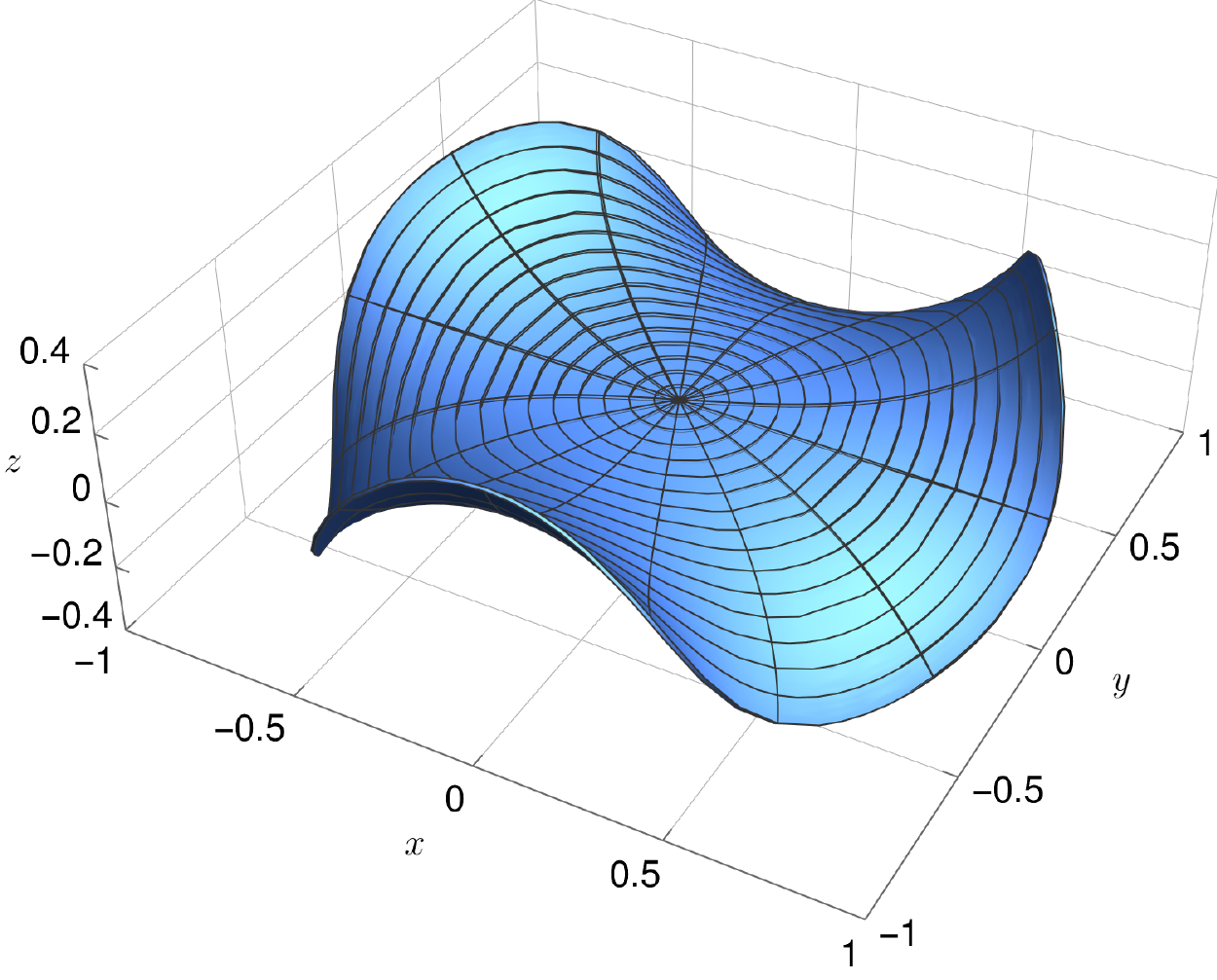}\\
\includegraphics[width=0.5\textwidth]{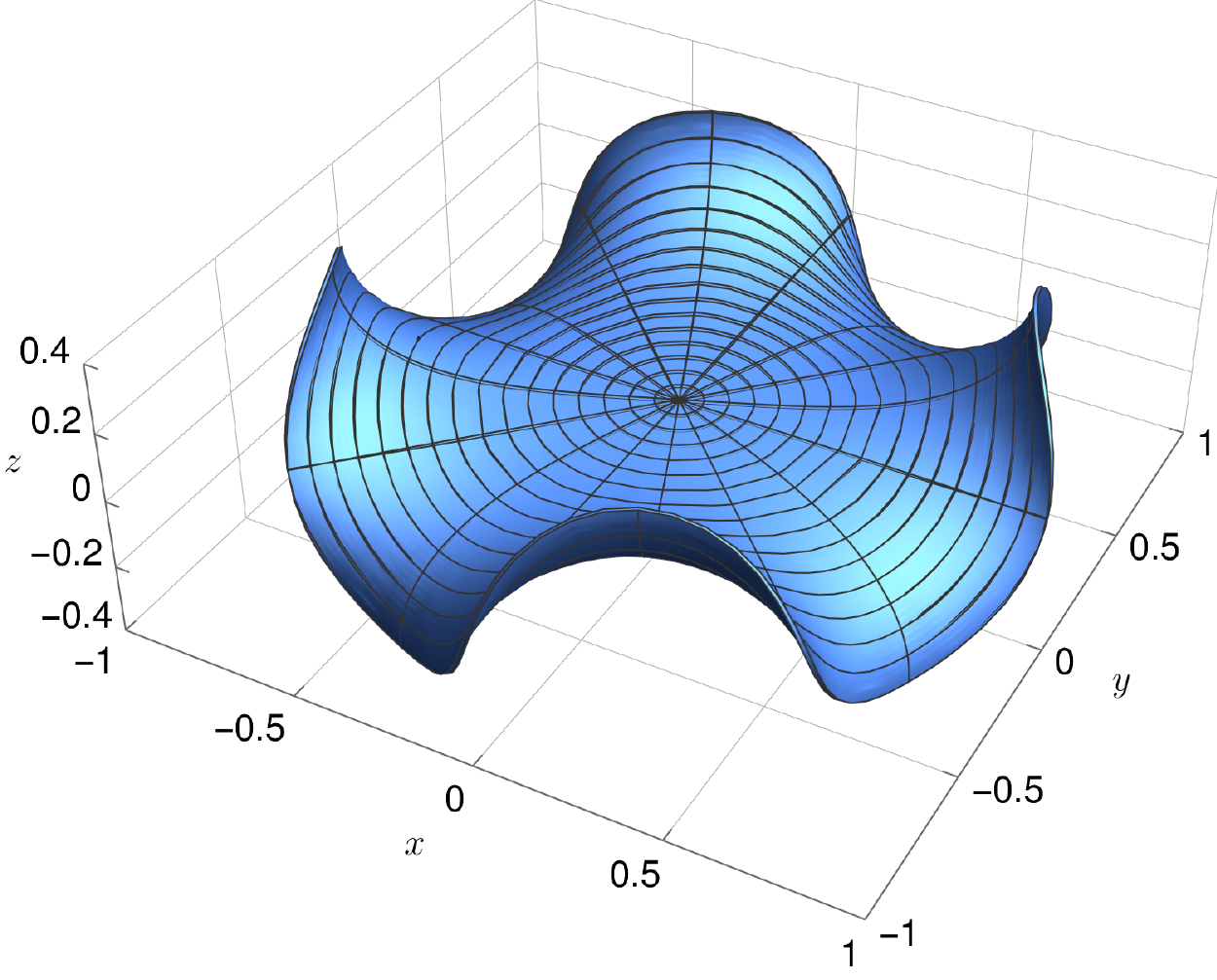}\includegraphics[width=0.5\textwidth]{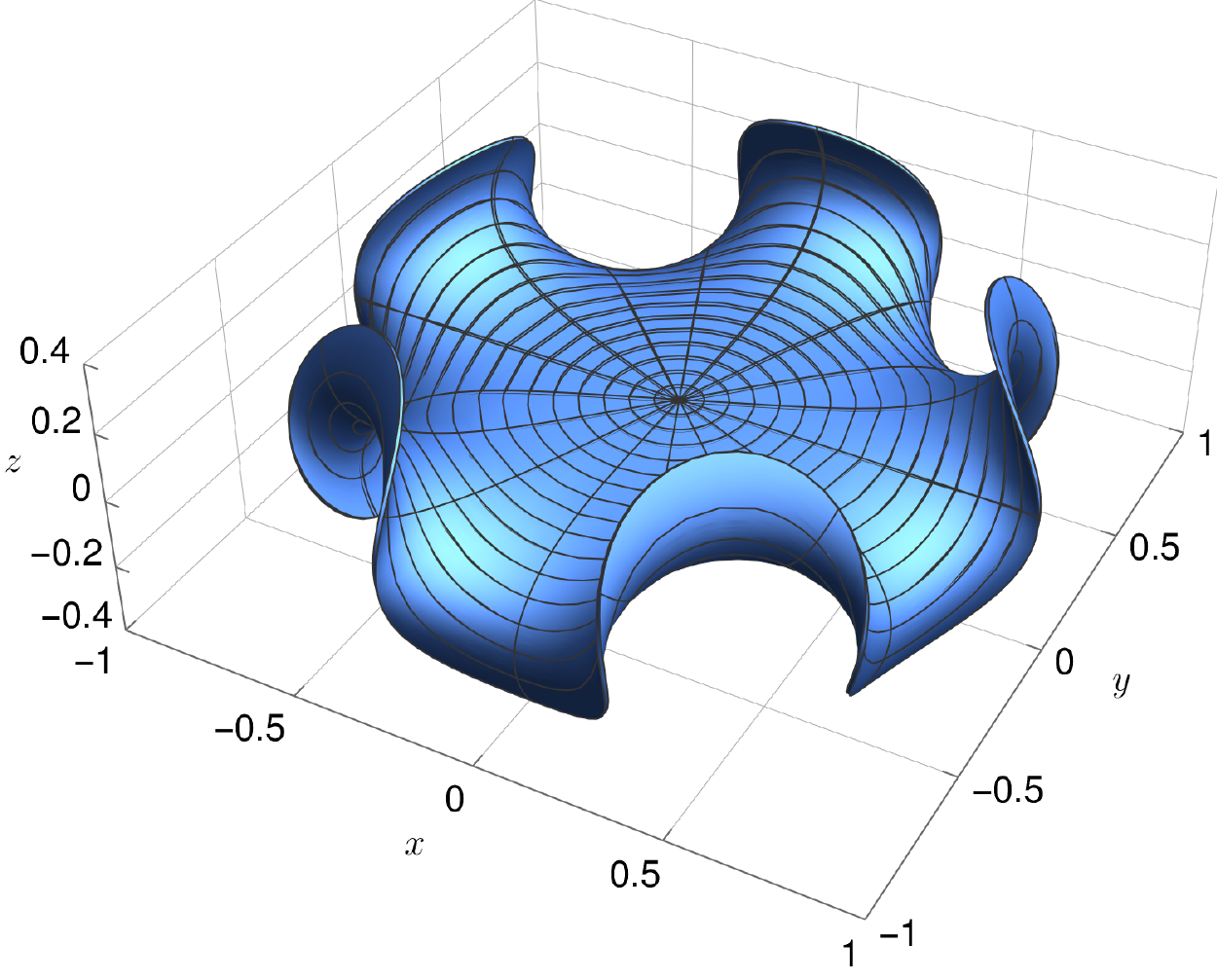}
\caption{Plot of actual buckled configuration of the elastic plate (using dimensional variables) as given by Eq.~\eqref{solnegcg} for the initial stress distribution in Eq.~\eqref{tau0neg}. We have set $H=0.01$ and $L=1$, where $S^m = \{\vect{Y}\in\mathbb{R}^2\;|\;\|\vect{Y}\|<1\}$. The other parameters are (top, left) $n=0$, $c=80$; (top, right), $n=1$, $c = 200$; (bottom, left), $n=2$, $c = 400$; (bottom, right), $n=3$, $c = 800$.}
\label{figcgneg}
\end{figure}
From Eq.~\eqref{fvk1},  the spontaneous Gaussian curvature is negative, being:
\begin{equation}
 C_G= -c^2 R^{2 n} <0,
\end{equation}
where we have a constant Gaussian curvature if $n=0$.
Seeking for a solution that cancels the second Piola-Kirchhoff  stress tensor, Eqs.~\eqref{fvk1}-\eqref{fvk2} become a Monge-Ampère and a biharmonic equation respectively, namely
\begin{equation}
 [\xi,\xi]+c^2 R^{2 n}= 0, \qquad   \Updelta_m^2 \xi=0.
\end{equation}
It is easy to check that these equations admit a solution for each $n\in\mathbb{N}$ of the form
 \begin{equation}
 \xi(R,\,\Theta) = \frac{c R^{n+2} \sin ((n+2) \Theta)}{n^2+3 n+2}.
 \label{solnegcg}
\end{equation}
In particular, the solution for $n=0$ corresponds to a twisting of the plate, as illustrated in Figure~\ref{figcgneg}.

\subsubsection{Concentrated Gaussian curvature in thin plates}

In this subsection, we consider the case of large deflections $\xi$, so that the bending term $\Updelta_m^2\xi$ in Eq.~\eqref{fvk2} can be neglected \cite{dervaux2008morphogenesis}. In this case, if $\chi=0$ the second FvK equation is automatically satisfied while the first FvK equation \eqref{fvk1} becomes the Monge-Ampère equation
\begin{equation}
[\xi,\xi] = \Updelta_m^2 \chi_0.
\end{equation}

\begin{figure}
\centering
\includegraphics[width=0.33\textwidth]{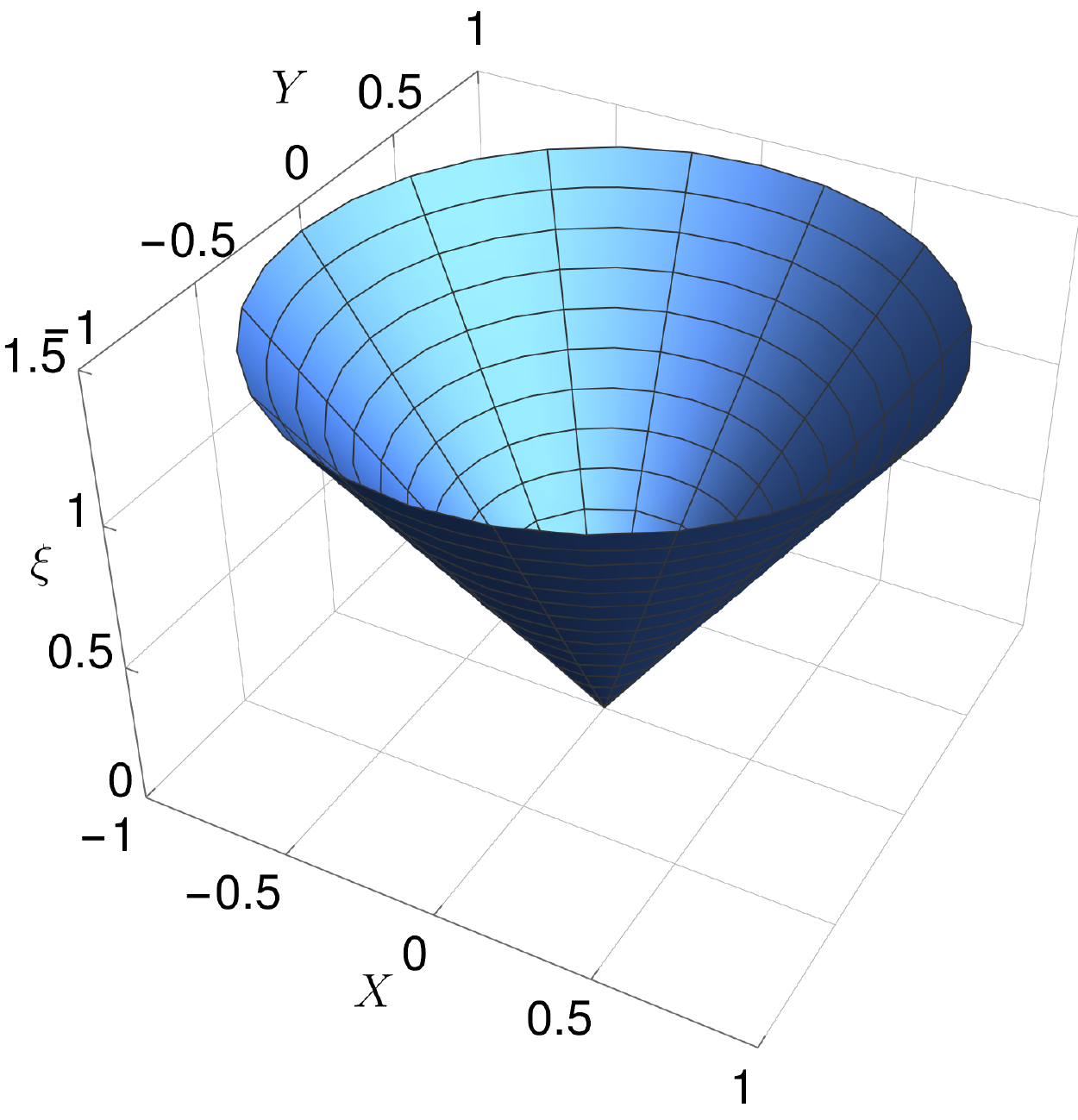}%
\includegraphics[width=0.33\textwidth]{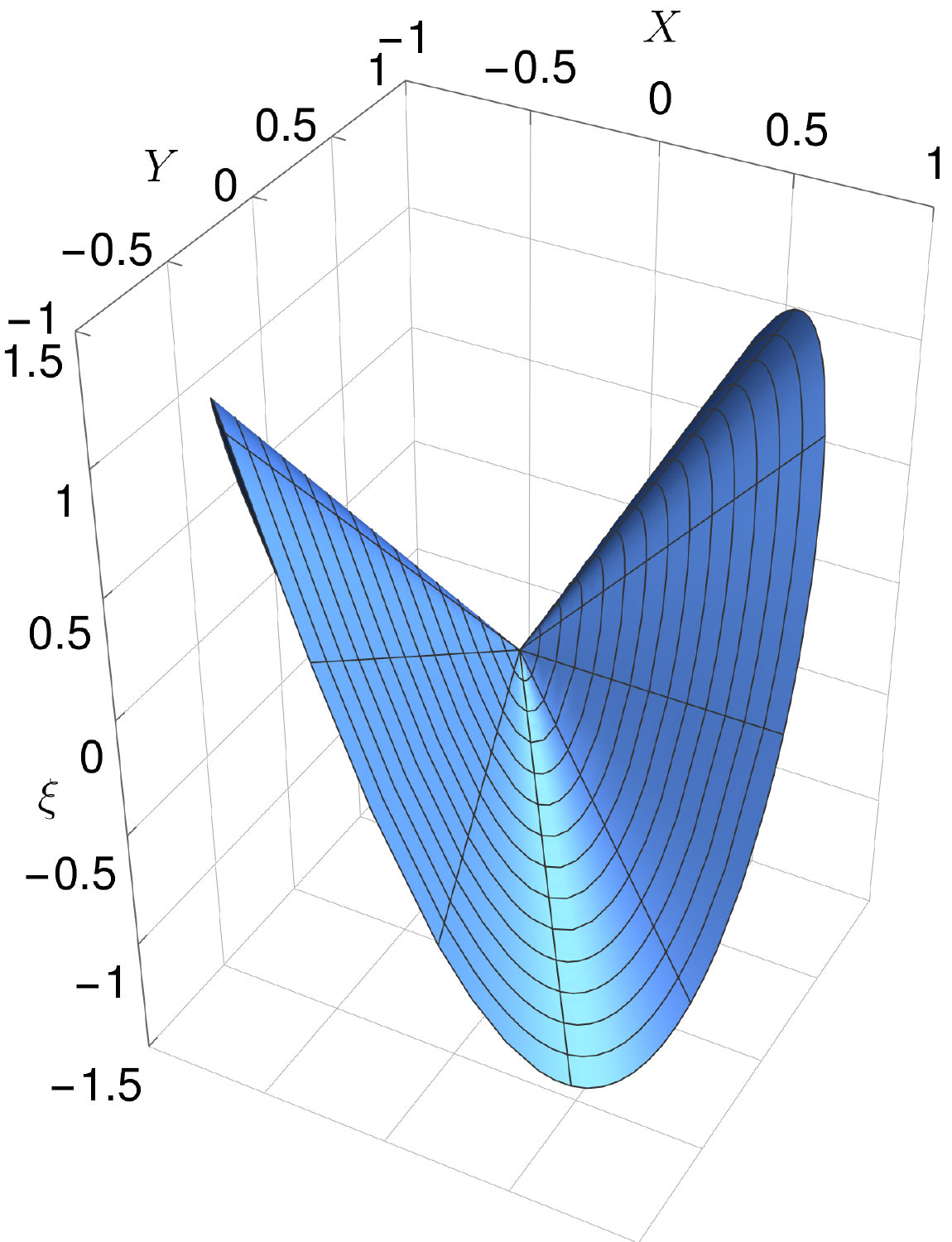}%
\includegraphics[width=0.33\textwidth]{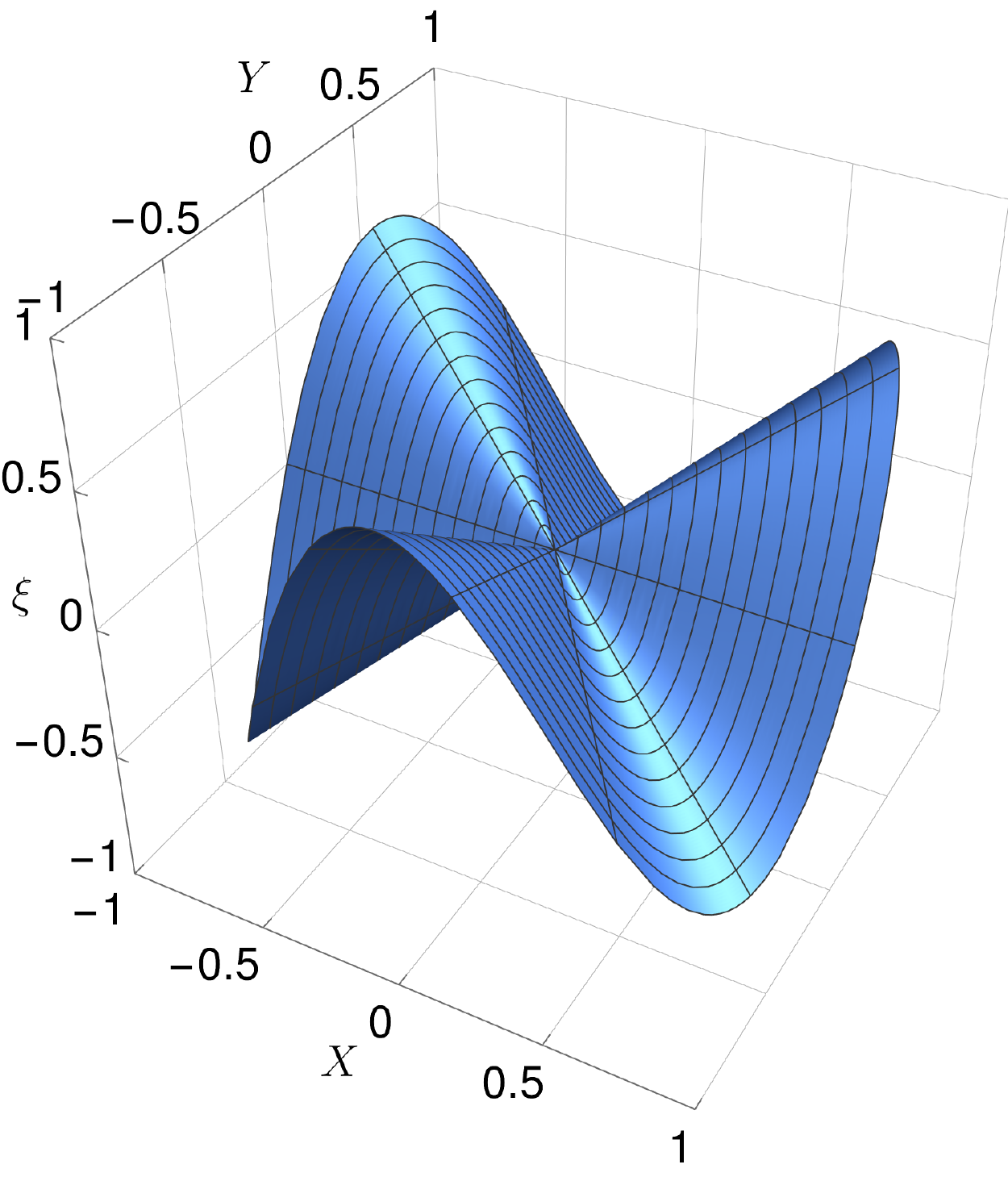}
\caption{Plot of the vertical deflection $\xi$ of the elastic plate as given by Eq.~\eqref{eq:sol_dirac} for the initial stress distribution in Eq.~\eqref{eq:logarithmic} where we set $\alpha=1$ (left) and $\alpha=-1$ (center and right). In the solutions for $\alpha=-1$ we fix $m=2$ (center) and $m=3$ (right).}
\label{fig:delta}
\end{figure}

We now consider the initial Airy stress function
\begin{equation}
\chi_0(R,\,\Theta)= \frac{\alpha R^2}{4}(\log R -1),
\end{equation}
which corresponds to the planar initial stress tensor
\begin{equation}
\label{eq:logarithmic}
\boldsymbol{\tau}_m = \frac{\alpha}{4}(2\log R + 1)\tens{I} - \frac{\alpha}{2}\vect{E}_R\otimes\vect{E}_R.
\end{equation}
We get that $\Updelta_m^2 \chi_0 = \alpha \delta(R)/R$, where $\delta(R)$ is a Dirac delta such that
\begin{equation}
\int_0^{+\infty} f(R)\delta(R) dR = f(0).
\end{equation}
If the second Piola-Kirchhoff stress tensors vanishes, the first FvK equation \eqref{fvk1} becomes
\begin{equation}
\label{eq:dirac}
-\frac{\xi _{,\Theta}^2}{R^4}+\frac{2 \xi _{,R\Theta} \xi _{,\Theta}}{R^3}+\frac{\xi _{,\Theta\Theta} \xi _{,RR}}{R^2}-\frac{\xi _{,R\Theta}^2}{R^2}+\frac{\xi _{,R} \xi _{,RR}}{R} = \alpha \frac{\delta(R)}{R},
\end{equation}
which admits as solutions the functions $\xi(R,\,\Theta) = g(\Theta) R$, where $g(\Theta)$ must satisfy the relation \cite{ben1997crumpled}
\begin{equation}
\int_0^{2\pi}g(\Theta)\left(g''(\Theta)+g(\Theta)\right)\,d\Theta = 4 \pi \alpha.
\end{equation}
Some solutions of Eq.~\eqref{eq:dirac} are given by \cite{xu2020water}
\begin{equation}
\label{eq:sol_dirac}
\begin{aligned}
\xi(R,\,\Theta) &= \sqrt{2 \alpha} R, &&\text{if }\alpha>0,\\
\xi(R,\,\Theta) &= 2R\sqrt{\frac{-\alpha}{m^2-1}}\sin(m\theta), &&\text{if }\alpha<0\text{, with }m\in\mathbb{N}\setminus\{0,\,1\}.
\end{aligned}
\end{equation}
The plots of some of these solutions are shown in Figure~\ref{fig:delta}.

\subsubsection{Inhomogeneous spontaneous Gaussian curvature}

Let us now consider an initial Airy stress function given by:
\begin{equation}
 \chi_0(X,\,Y)= -c^2 \ e^{-2 k \ X},
\end{equation}
with $c$ being a characteristic dimensionless stress parameter and $k>0$ a characteristic dimensionless decay length. It corresponds to a uniaxial compression along the $Y$ axis  that is inhomogeneous along the $X$ axis, and  the only non vanishing initial stress component is:
 \begin{equation}
 \tau^m_{YY}= -4 k^2 c^2 \ e^{-2 k  X}.
\label{tau0inh}
\end{equation}
From Eq.~\eqref{fvk1},  the spontaneous Gaussian curvature is also inhomogeneous, being:
\begin{equation}
 C_G= -16 c^2 k^4 e^{-2 k  X}.
\end{equation}
This inhomogeneous expression for $C_G$ was obtained  in \cite{dervaux2009morphogenesis} using a different approach, based on imposing an incompatible distortion of the elastic metric. Through the cancellation of the second Piola-Kirchhoff  stress tensor in the actual configuration, Eqs.~\eqref{fvk1}-\eqref{fvk2} simplify as the following Monge-Ampère and biharmonic equations:
 \begin{equation}
 [\xi,\xi] + 16 c^2 k^4 e^{-2 k X}= 0, \qquad   \Updelta_m^2 \xi=0.
 \label{fvkinhcg}
\end{equation}

\begin{figure}[!t]
\centering
\includegraphics[width=0.5\textwidth]{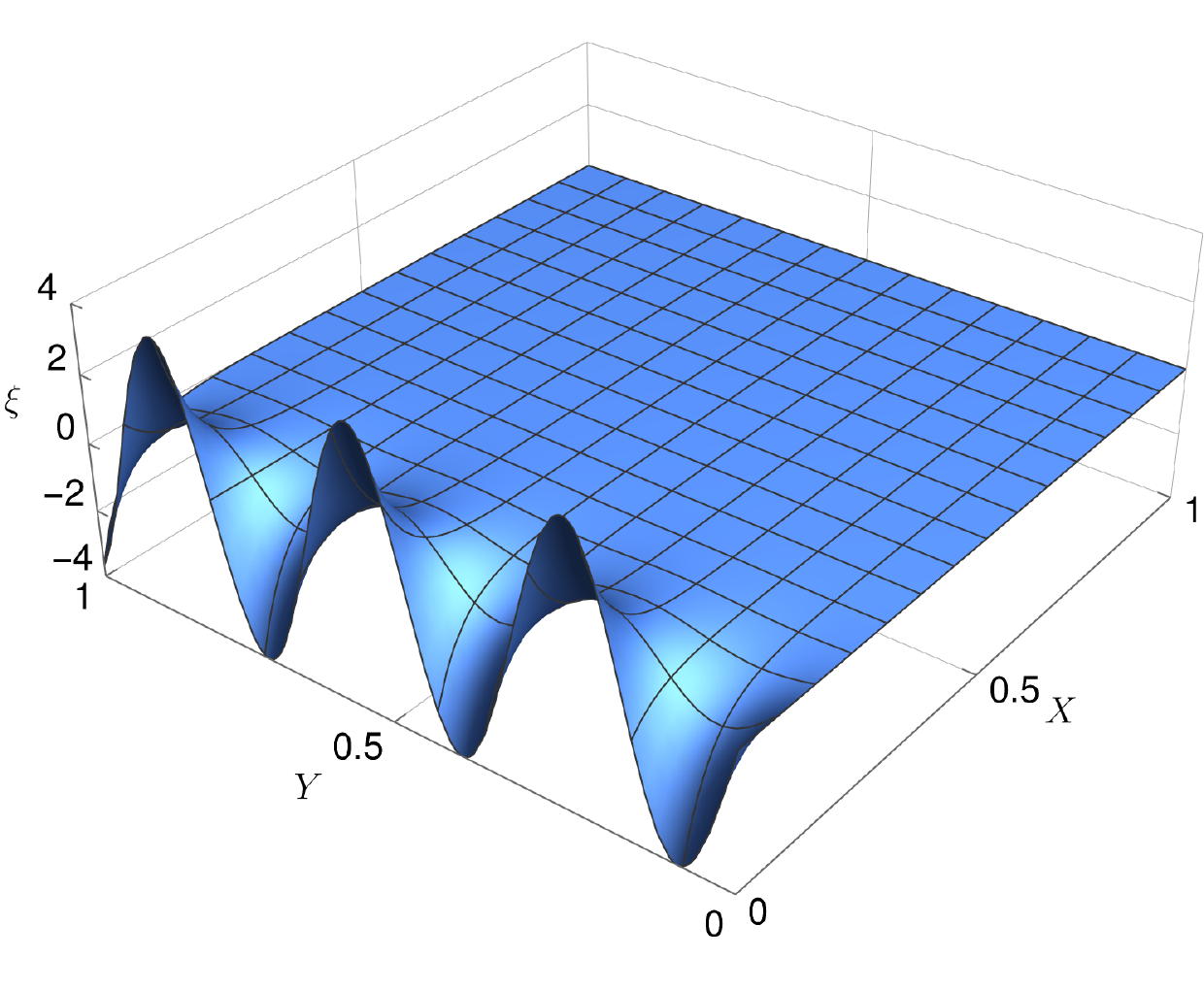}
\caption{Plot of the vertical deflection $\xi$ of the elastic plate as given by Eq.~\eqref{solinhcg} for the initial stress distribution in Eq.~\eqref{tau0inh} where we set $c=1$ and $k=20$. }
\label{figcginh}
\end{figure}

Assuming variable separation, we set $\xi= f(Y) \ e^{- k  X}$ so that Eq.~\eqref{fvkinhcg} can be transformed in the following nonlinear ordinary differential system:
 \begin{equation}
 16 c^2 k^2 - f'^2 + f \ f''=0, \quad k^4 f + 2 k^2 f'' + f''''=0.
 \label{fvkinhcg2}
\end{equation}
A solution of Eq.~\eqref{fvkinhcg2} is given by $f= 4 c \sin (k  Y)$, so that the vertical deflection is :
 \begin{equation}
 \xi=  4 c \sin (k  Y) \  e^{- k X},
 \label{solinhcg}
\end{equation}
that corresponds  to a sinusoidal oscillation along the $Y$ axis that decays exponentially along the   $X$ axis with a decay length $k$ and an amplitude proportional to $c$. The resulting morphology of the middle plane of the initially stressed plate is depicted in Figure~\ref{figcginh}.

\subsection{Initial stress varying along the plate thickness}

We finally consider the case in which the initial stress is allowed to vary along the plate thickness. For the sake of simplicity,  we take ${\boldsymbol \tau}= {\boldsymbol \tau (X,Z)}$, with $ \tau_{YY}= \tau_{XY}=\tau_{YZ}=0$. In this case, from Eq.~\eqref{atau2} we can define an Airy stress function $\Phi= \Phi(X,Z)$ such that:
\begin{equation}
 \tau_{XX}= \Phi_{,ZZ}, \quad  \tau_{XZ}= -\Phi_{,XZ}, \quad \tau_{ZZ}= \Phi_{,XX}. 
\end{equation}
In particular, assuming that  $\tau_{XZ}|_{Z\pm 1}= \tau_{ZZ}|_{Z\pm 1}=0$, a leading order expansion of $\Phi$ along the vertical direction gives:
\begin{equation}
 \Phi= \left(\frac{Z^3}{3} - Z\right )(a_1 X+ a_0) +  b \ Z^2,
\end{equation}
where $a_0$,$a_1$, and $b$ are dimensionless stress parameters. The only non zero-components of the average  initial stress tensor ${\boldsymbol \tau}_m$ and torque tensor $\tens{M}$ are the following:
\begin{equation}
 \tau^m_{XX}= 2b,  \quad \tau^m_{XZ} = \frac{2}{3}a_1, \quad M_{XX}= \frac{2}{3}(a_1 X+ a_0).
\label{taum}
\end{equation}

\begin{figure}[!b]
\centering
\includegraphics[width=0.5\textwidth]{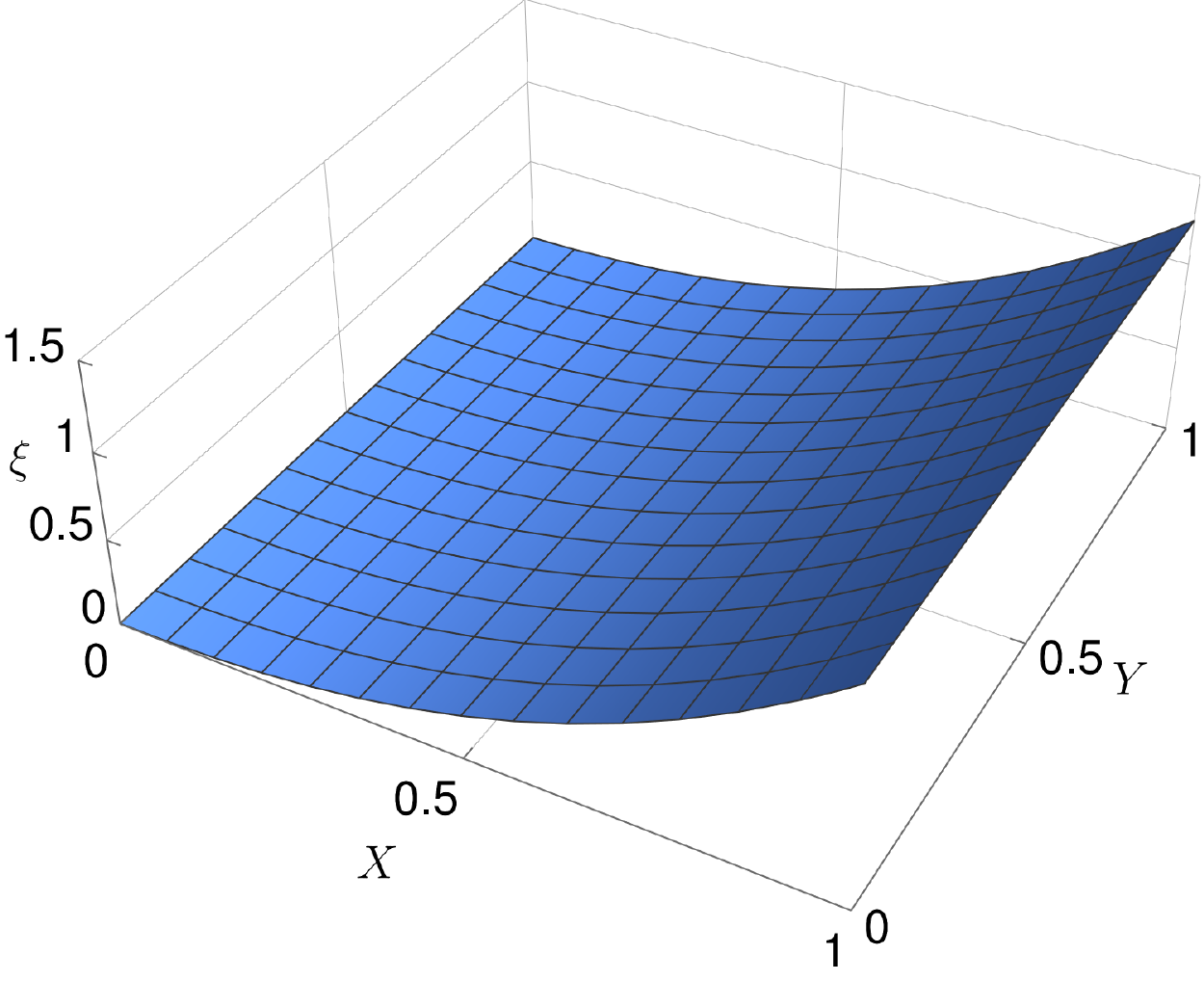}\includegraphics[width=0.5\textwidth]{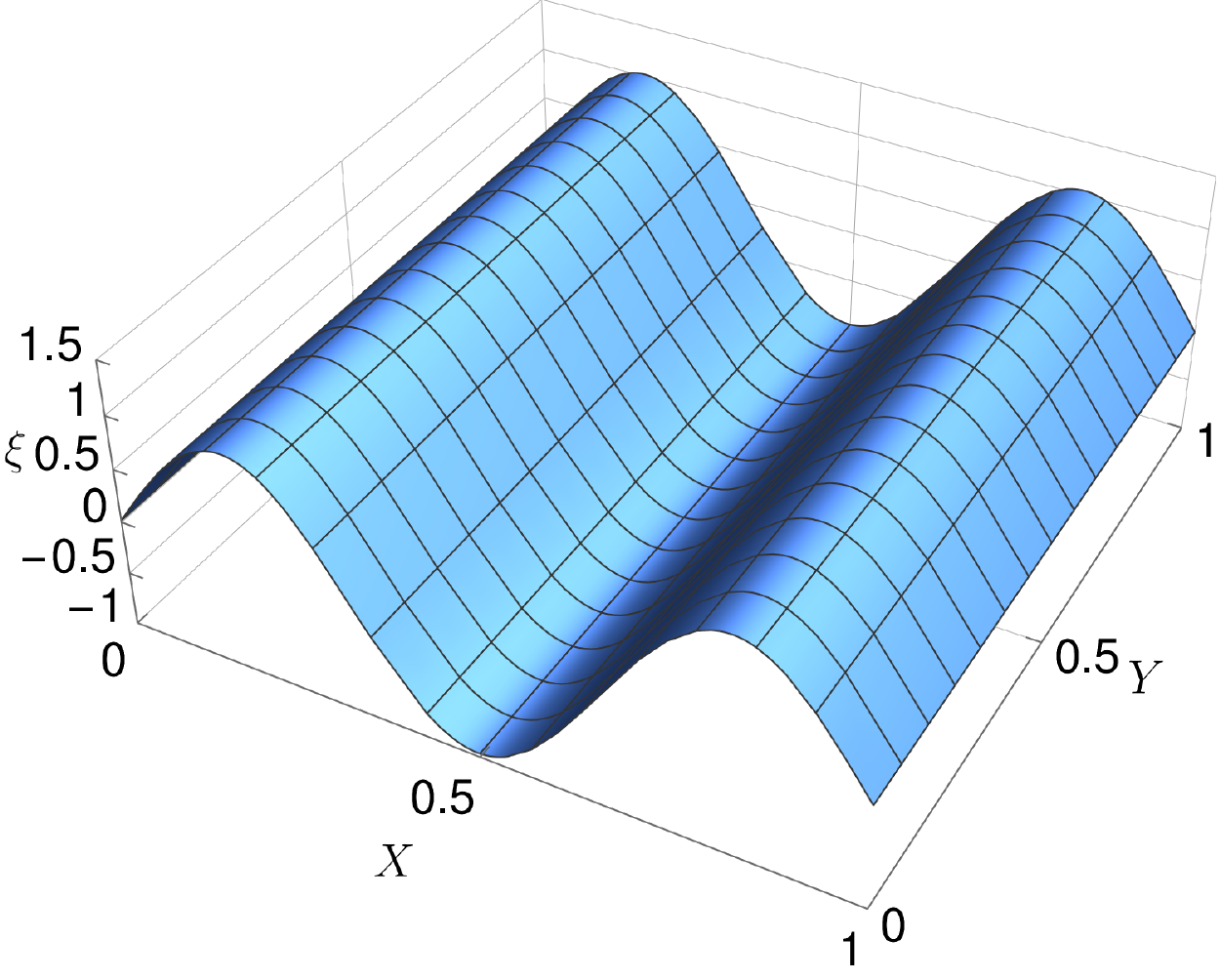}
\caption{Plot of the vertical deflection $\xi$ of the elastic plate, with the initial stress distribution given by Eq.~\eqref{taum} setting $a_0=1$, $a_1=2$.  (Left) Spontaneous bending with zero average initial stress $\tau^m_{XX}$, setting $b=0$ and $C_1=C_2=0$ in Eq.~\eqref{sol1}.  (Right) Asymmetric buckling with negative average initial stress $\tau^m_{XX}$, setting  $b=-20$ and $C_1=-i$, $C_2=0$ for the initial stress distribution in Eq.~\eqref{sol2}. }
\label{figbend}
\end{figure}    

Accordingly, the parameter $b$ determines the characteristic amplitude of the average initial stress, while $a_0$,$a_1$ determine the magnitude of the initial bending torque. Since $\tau^m_{XX}$ is a constant, then the initial stress distribution has zero spontaneous Gaussian curvature, i.e. $C_G=0$. Assuming $\xi=\xi(X)$,  since $[\xi,\xi]=0$ Eq.~\eqref{fvk1} is automatically satisfied if $\chi=\chi_0= b Y^2$. The torque distribution in Eq.~\eqref{taum} imposes a spontaneous average curvature $C_M=\frac{3}{2}(a_1 X+ a_0)$, and Eq.~\eqref{fvk2} reads:
\begin{equation}
\frac{2}{9}\xi^{''}-{b} \xi -\frac{1}{3}(a_1 X+ a_0)=0.
\label{fvk2t}
\end{equation}
If the average initial stress is zero, i.e. if $b=0$, then the solution corresponds to a bending deflection having exactly the  curvature $C_M$, with:
\begin{equation}
\xi= \frac{3}{4} \left(a_0 X^2 - (a_1 X^3)/3\right) + C_1 + C_2 X,
\label{sol1}
\end{equation}
where $C_1$, $C_2$ are two constants of integration that must be fixed by boundary conditions.
  
If the average initial stress does not vanish, i.e. $b\ne0$,  the deflection reads:
\begin{equation}
\xi= -\frac{a_0 +a_1X}{3 b} + C_1 \sinh\left(\frac{3X\sqrt{b}}{\sqrt{2}}\right) + C_2 \cosh\left(\frac{3{X\sqrt{b}}}{\sqrt{2}}\right).
\label{sol2}
\end{equation}
If the average uniaxial stress is tensile, i.e. $b>0$, than the deflection has an exponential trend; if it is compressive , i.e. $b<0$, than the deflection has the characteristic  sinusoidal behavior of an Euler buckling where the symmetry is broken by a spontaneous average curvature. 
Two possible bending morphologies of the middle line of the plate are depicted in Figure~\ref{figbend}.

Thus, in the limit of narrow plates, the governing equation Eq.~\eqref{fvk2t} recovers  the Euler beam theory for elastic rods, where the average initial stress $\tau^m_{XX}$ appears as a traction load, and  $M_{XX}$ as a distributed torque provoking a spontaneous curvature

\section{Concluding remarks}

In this work, we have derived the FvK equations for an elastic plate with initial stress. The reference configuration of the plate is a parallelepiped, whose thickness $2H$ along the $z$-direction is much smaller than the characteristic length  $L$ of its edges. We have identified with $\epsilon=H/L$ the dimensionless parameter describing the thinness of the plate.
Adopting the theoretical framework of initially stressed materials developed in \cite{shams}, the elastic energy of the plate is obtained by an asymptotic expansion  with respect to the small parameter $\epsilon$.

More explicitly, we have assumed that the plate is composed of an incompressible neo-Hookean material, whose strain energy depends on both the deformation gradient and the initial stress, see \eqref{psinh}. 
Assuming the classical Kirchhoff hypothesis for the displacement of the plate (see \eqref{eq:u_parallel}-\eqref{W}), we used the scaling for the initial stress components reported in  \eqref{atau}. 


Under these assumptions, we have obtained the balance equation through a variational approach, by enforcing that the energy functional be stationary. 
The equilibrium equations \eqref{fvk1} and \eqref{fvk2} generalize the FvK equations for the average planar and bending stretch, respectively, to the case of initially stressed plates. 

We have solved the FvK equations in some physical examples of engineering interest. 
By tuning the initial stress distribution within the plate, we have obtained buckled configurations exhibiting a constant positive or negative Gaussian curvature. 
Furthermore, it is possible to obtain surfaces with non constant Gaussian curvature by properly tuning the initial stress of the plate. 
In particular, we propose some new solutions with a radially inhomogeneous Gaussian curvature. We have also recovered some known explicit solutions of the FvK equations, such as the conical solutions proposed by \cite{ben1997crumpled} for thin plates, where the initial stress exhibits logarithmic singularities, and the solution for a geometrically frustrated FvK plate reported in \cite{dervaux2009morphogenesis}.
Finally, we analyzed the effect of an initial stress that varies along the plate thickness. 
The plate spontaneously bend when the initial stress along the $X$-direction is non negative, while it exhibits a wrinkling pattern if it is compressive, similarly to Euler buckling of elastic beams. 
In the case of more complex initial stress distributions, e.g. exhibiting checkerboard or labyrinth patterns, it is unlikely that analytical solutions can be found and approximated solutions of the proposed equations may be found exploiting numerical techniques, such as the finite element method.

In conclusion, we have derived the FvK equations for an initially stressed plate using a formal dimensional reduction of a nonlinear strain energy function depending explicitly on both the deformation gradient and the initial stress tensor. The main advance of the proposed approach  is to unravel the effects oh the initial stress distribution on the spontaneous average and Gaussian curvatures of the plate without the need to prescribe  incompatible pre-strains, as required in earlier works \cite{klein2007shaping,dervaux2009morphogenesis}.  In fact, unlike pre-straining, the initial stress distribution  within the body can be measured by means of  non-destructive techniques, such as ultrasound elastography \cite{li2017guided} or photoelasticity \cite{nienhaus2009determination}. Moreover, a target initial stress distribution can be physically realized using  modern digital fabrication techniques,  such as 4D printing \cite{zurlo2017printing,van2018mechanics} and UV lithography  \cite{na2016grayscale}. Thus, the results of the proposed model may be used to design residually stressed object in proximity of an elastic bifurcation \cite{Jones_2015}, with the aim to fabricate shape-shifting plates able to adapt their morphology in presence of external stimuli, such as a chemical potential \cite{andrini2021theoretical} or an electric field \cite{de2010pull}. 

\subsubsection*{Funding} This work was partly supported by MUR, PRIN  Research Projects 2017KL4EF3 and 2020F3NCPX, by the National
Group of Mathematical Physics (GNFM -- INdAM) through
the program \emph{Progetto Giovani 2020}, and  by Regione Lombardia project NEWMED (Grant No. POR FESR 2014-2020).

\appendix
\section{Expression of the tensor $\tens{C}$ and of the F\"{o}ppl--von K\'{a}rm\'{a}n equations in Cartesian coordinates}
\label{appendix}
In this appendix we express the right Cauchy-Green tensor $\tens{C}$ and the FvK equations using a Cartesian coordinate system. In the following, all the quantities are dimensionless.

Let $U(X,Y)$ and $V(X,\,Y)$ be the components of $\vect{U}(X,\,Y)$ with respect to the basis $\vect{E}_X$ and $\vect{E}_Y$, namely
\begin{equation}
\vect{U}(X,\,Y) = U(X,\,Y)\vect{E}_X + V(X,\,Y)\vect{E}_Y.
\end{equation}
The right Cauchy-Green tensor reads $\tens{C} = \tens{I} + \epsilon^2 \tens{A}+ o(\epsilon^2)$, where $\tens{A} = \tens{C}_0 + Z \tens{C}_1$ and
\begin{equation}
\begin{gathered}
\tens{C}_0 = 
 \begin{bmatrix}
 \xi_{,X}^2+2 U_{,X}& \xi_{,X}\xi_{,Y}+U_{,Y}+V_{,X} & 0 \\
 \xi_{,X}\xi_{,Y} +U_{,Y}+V_{,X} & \xi_{,Y}^2+2 V_{,Y} & 0 \\
 0 & 0 & -\|\nabla\xi\|^2-2 U_{,X}-2V_{,Y}
\end{bmatrix},\\
\tens{C}_1 =
\begin{bmatrix}
 -2 \xi_{,XX} & -2 \xi_{,XY} & 0 \\
 -2 \xi_{,XY} & -2 \xi_{,YY} & 0 \\
 0 & 0 & 2 \Updelta\xi\\
\end{bmatrix}
\end{gathered}.
\end{equation}

Finally, in order to retrieve the FvK equations using the canonical vector basis $(\vect{E}_X,\,\vect{E}_Y,\,\vect{E}_Z)$,  we recall that
\begin{equation}
[a,b]= \frac{1}{2}(a_{,XX} b_{,YY}+a_{,YY} b_{,XX}-2 a_{,XY} b_{,XY}).
\end{equation}
so that Eq.~\eqref{fvk1} becomes
\begin{equation}
\begin{multlined}
\xi _{,XX}\xi _{,YY} -\xi_{,XY}^2+\chi _{,XXXX}+2 \chi _{,XXYY}+\chi _{,YYYY}+\\
-{\chi_0}_{,XXXX}-2 {\chi_0}_{,XXYY}-{\chi_0}_{,YYYY}=0,
\end{multlined}
\end{equation}
while Eq.~\eqref{fvk2} is given by
\begin{equation}
\begin{multlined}
\frac{2}{9}\left( \xi _{,XXXX}+2 \xi _{,XXYY}+ \xi _{,YYYY}\right)-
\frac{1}{2} \left(M_{XX,XX}+2 M_{XY,XY}+ M_{YY,YY}\right)+\\
+\frac{1}{2}\left(2 \xi_{,XY} \chi_{,XY}- \xi _{,XX} \chi _{,YY}- \xi _{,YY} \chi _{,XX}\right)=0.
\end{multlined}
\end{equation}

\bibliographystyle{abbrv}

\begin{thebibliography}{99}

\bibitem{foppl1921vorlesungen}
F{\"o}ppl A. 1921 {\em Vorlesungen {\"u}ber technische Mechanik} vol.~6.
BG Teubner.

\bibitem{karman1907festigkeitsprobleme}
von K{\'a}rm{\'a}n T. 1907  Festigkeitsprobleme im maschinenbau. In {\em
  Mechanik} ,  pp. 311--385. Springer.

\bibitem{goodier1938problems}
Goodier JN. 1938  On the problems of the beam and the plate in the theory of
  elasticity. Royal Society of Canada.

\bibitem{friedrichs1961boundary}
Friedrichs K, Dressler R. 1961  A boundary-layer theory for elastic plates.
  {\em Communications on Pure and Applied Mathematics} \textbf{14}, 1--33.

\bibitem{gol1962derivation}
Gol'denveizer A. 1962  Derivation of an approximate theory of bending of a
  plate by the method of asymptotic integration of the equations of the theory
  of elasticity. {\em Journal of Applied Mathematics and Mechanics}
  \textbf{26}, 1000--1025.

\bibitem{ciarlet1980justification}
Ciarlet PG. 1980  {A justification of the von K{\'a}rm{\'a}n equations}. {\em
  Archive for Rational Mechanics and Analysis} \textbf{73}, 349--389.

\bibitem{ciarlet2006equations}
Ciarlet PG, Rabier P. 2006 {\em Les {\'e}quations de von K{\'a}rm{\'a}n} vol.
  826.
Springer.

\bibitem{ben1997crumpled}
Ben~Amar M, Pomeau Y. 1997  Crumpled paper. {\em Proceedings of the Royal
  Society of London. Series A: Mathematical, Physical and Engineering Sciences}
  \textbf{453}, 729--755.

\bibitem{cerda2003geometry}
Cerda E, Mahadevan L. 2003  Geometry and physics of wrinkling. {\em Physical
  Review Letters} \textbf{90}, 074302.

\bibitem{audoly2008buckling}
Audoly B, Boudaoud A. 2008  {Buckling of a stiff film bound to a compliant
  substrate—Part I:: Formulation, linear stability of cylindrical patterns,
  secondary bifurcations}. {\em Journal of the Mechanics and Physics of Solids}
  \textbf{56}, 2401--2421.

\bibitem{audoly2010elasticity}
Audoly B, Pomeau Y. 2010 {\em Elasticity and Geometry: From Hair Curls to the
  Non-linear Response of Shells}.
Oxford University Press.

\bibitem{vella2017indentation}
Vella D, Davidovitch B. 2017  Indentation metrology of clamped, ultra-thin
  elastic sheets. {\em Soft Matter} \textbf{13}, 2264--2278.

\bibitem{fernandes2010multiparameter}
Fernandes A, Maurini C, Vidoli S. 2010  Multiparameter actuation for shape
  control of bistable composite plates. {\em International Journal of Solids
  and Structures} \textbf{47}, 1449--1458.

\bibitem{felton2014method}
Felton S, Tolley M, Demaine E, Rus D, Wood R. 2014  A method for building
  self-folding machines. {\em Science} \textbf{345}, 644--646.

\bibitem{zhang2014defects}
Zhang T, Li X, Gao H. 2014  Defects controlled wrinkling and topological design
  in graphene. {\em Journal of the Mechanics and Physics of Solids}
  \textbf{67}, 2--13.

\bibitem{gladman2016biomimetic}
Gladman AS, Matsumoto EA, Nuzzo RG, Mahadevan L, Lewis JA. 2016  Biomimetic 4D
  printing. {\em Nature Materials} \textbf{15}, 413--418.

\bibitem{bowick2017non}
Bowick MJ, Ko{\v{s}}mrlj A, Nelson DR, Sknepnek R. 2017  Non-Hookean
  statistical mechanics of clamped graphene ribbons. {\em Physical Review B}
  \textbf{95}, 104109.

\bibitem{van2018mechanics}
van Rees WM, Matsumoto EA, Gladman AS, Lewis JA, Mahadevan L. 2018  Mechanics
  of biomimetic 4D printed structures. {\em Soft Matter} \textbf{14},
  8771--8779.

\bibitem{tao2021morphing}
Tao Y, Lee YC, Liu H, Zhang X, Cui J, Mondoa C, Babaei M, Santillan J, Wang G,
  Luo D et~al.. 2021  Morphing pasta and beyond. {\em Science Advances}
  \textbf{7}, eabf4098.

\bibitem{feinberg2007muscular}
Feinberg AW, Feigel A, Shevkoplyas SS, Sheehy S, Whitesides GM, Parker KK. 2007
   Muscular thin films for building actuators and powering devices. {\em
  Science} \textbf{317}, 1366--1370.

\bibitem{hines2017soft}
Hines L, Petersen K, Lum GZ, Sitti M. 2017  Soft actuators for small-scale
  robotics. {\em Advanced materials} \textbf{29}, 1603483.

\bibitem{kim2010dynamic}
Kim J, Yoon J, Hayward RC. 2010  Dynamic display of biomolecular patterns
  through an elastic creasing instability of stimuli-responsive hydrogels. {\em
  Nature Materials} \textbf{9}, 159--164.

\bibitem{pezzulla2015morphing}
Pezzulla M, Shillig SA, Nardinocchi P, Holmes DP. 2015  Morphing of geometric
  composites via residual swelling. {\em Soft Matter} \textbf{11}, 5812--5820.

\bibitem{van2017growth}
van Rees WM, Vouga E, Mahadevan L. 2017  Growth patterns for shape-shifting
  elastic bilayers. {\em Proceedings of the National Academy of Sciences}
  \textbf{114}, 11597--11602.

\bibitem{zurlo2017printing}
Zurlo G, Truskinovsky L. 2017  {Printing non-Euclidean solids}. {\em Physical
  Review Letters} \textbf{119}, 048001.

\bibitem{Mihai_2020}
Mihai LA, Goriely A. 2020  A plate theory for nematic liquid crystalline
  solids. {\em Journal of the Mechanics and Physics of Solids} \textbf{144},
  104101.

\bibitem{yue2017modified}
Yue Y, Ru C, Xu K. 2017  Modified von K{\'a}rm{\'a}n equations for elastic
  nanoplates with surface tension and surface elasticity. {\em International
  Journal of Non-Linear Mechanics} \textbf{88}, 67--73.

\bibitem{dervaux2009morphogenesis}
Dervaux J, Ciarletta P, Ben~Amar M. 2009  Morphogenesis of thin hyperelastic
  plates: a constitutive theory of biological growth in the F{\"o}ppl--von
  K{\'a}rm{\'a}n limit. {\em Journal of the Mechanics and Physics of Solids}
  \textbf{57}, 458--471.

\bibitem{Friesecke_2006}
Friesecke G, James RD, Müller S. 2006  A Hierarchy of Plate Models Derived
  from Nonlinear Elasticity by Gamma-Convergence. {\em Archive for Rational
  Mechanics and Analysis} \textbf{180}, 183--236.

\bibitem{neukamm2013derivation}
Neukamm S, Vel{\v{c}}i{\'c} I. 2013  Derivation of a homogenized von-Karman
  plate theory from 3D nonlinear elasticity. {\em Mathematical Models and
  Methods in Applied Sciences} \textbf{23}, 2701--2748.

\bibitem{lewicka2011foppl}
Lewicka M, Mahadevan L, Pakzad MR. 2011  The F{\"o}ppl-von K{\'a}rm{\'a}n
  equations for plates with incompatible strains. {\em Proceedings of the Royal
  Society A: Mathematical, Physical and Engineering Sciences} \textbf{467},
  402--426.

\bibitem{Bhattacharya_2016}
Bhattacharya K, Lewicka M, Schäffner M. 2016  Plates with Incompatible
  Prestrain. {\em Archive for Rational Mechanics and Analysis} \textbf{221},
  143--181.

\bibitem{Ricciotti_2017}
Ricciotti D, Lewicka M, Raoult A. 2017  Plates with incompatible prestrain of
  high order. {\em Annales de l{\textquotesingle}Institut Henri Poincar{\'{e}}
  C, Analyse non lin{\'{e}}aire} \textbf{34}, 1883--1912.

\bibitem{Bola_os_2020}
Bola{\~{n}}os SJ. 2020  On the Dimension Reduction in Prestrained Elasticity
  (Survey). In {\em Advances in Mathematical Sciences} ,  pp. 311--324.
  Springer International Publishing.

\bibitem{klein2007shaping}
Klein Y, Efrati E, Sharon E. 2007  {Shaping of elastic sheets by prescription
  of non-Euclidean metrics}. {\em Science} \textbf{315}, 1116--1120.

\bibitem{dervaux2008morphogenesis}
Dervaux J, Ben~Amar M. 2008  Morphogenesis of growing soft tissues. {\em
  Physical Review Letters} \textbf{101}, 068101.

\bibitem{efrati2009elastic}
Efrati E, Sharon E, Kupferman R. 2009  {Elastic theory of unconstrained
  non-Euclidean plates}. {\em Journal of the Mechanics and Physics of Solids}
  \textbf{57}, 762--775.

\bibitem{efrati2013metric}
Efrati E, Sharon E, Kupferman R. 2013  The metric description of elasticity in
  residually stressed soft materials. {\em Soft Matter} \textbf{9}, 8187--8197.

\bibitem{ciarletta2016residual}
Ciarletta P, Destrade M, Gower AL. 2016  On residual stresses and homeostasis:
  an elastic theory of functional adaptation in living matter. {\em Scientific
  reports} \textbf{6}, 1--8.

\bibitem{hoger85}
Hoger A. 1985  On the residual stress possible in an elastic body with material
  symmetry. {\em Archive for Rational Mechanics and Analysis} \textbf{88},
  271--289.

\bibitem{shams}
Shams M, Destrade M, Ogden RW. 2011  Initial stresses in elastic solids:
  Constitutive laws and acoustoelasticity. {\em Wave Motion} \textbf{48},
  552--567.

\bibitem{merodio13}
Merodio J, Ogden RW, Rodr{\'\i}guez J. 2013  The influence of residual stress
  on finite deformation elastic response. {\em International Journal of
  Non-Linear Mechanics} \textbf{56}, 43--49.

\bibitem{spencer1971part}
Spencer A. 1971  {Part III. Theory of invariants}. In Eringen A, editor, {\em
  Continuum Physics, vol. 1} ,  pp. 239--353. Oxford: Academic Press.

\bibitem{gower15}
Gower AL, Ciarletta P, Destrade M. 2015  Initial stress symmetry and its
  applications in elasticity. {\em Proceedings of the Royal Society A:
  Mathematical, Physical and Engineering Sciences} \textbf{471}, 20150448.

\bibitem{Riccobelli_2019}
Riccobelli D, Agosti A, Ciarletta P. 2019  On the existence of elastic
  minimizers for initially stressed materials. {\em Philosophical Transactions
  of the Royal Society A: Mathematical, Physical and Engineering Sciences}
  \textbf{377}, 20180074.

\bibitem{schwartz1957theorie}
Schwartz L. 1957  Th{\'e}orie des distributions {\`a} valeurs vectorielles. I.
  In {\em Annales de l'institut Fourier} vol.~7 pp. 1--141.

\bibitem{xu2020water}
Xu F, Fu C, Yang Y. 2020  Water affects morphogenesis of growing aquatic plant
  leaves. {\em Physical review letters} \textbf{124}, 038003.

\bibitem{li2017guided}
Li GY, He Q, Mangan R, Xu G, Mo C, Luo J, Destrade M, Cao Y. 2017  Guided waves
  in pre-stressed hyperelastic plates and tubes: Application to the ultrasound
  elastography of thin-walled soft materials. {\em Journal of the Mechanics and
  Physics of Solids} \textbf{102}, 67--79.

\bibitem{nienhaus2009determination}
Nienhaus U, Aegerter-Wilmsen T, Aegerter CM. 2009  Determination of mechanical
  stress distribution in Drosophila wing discs using photoelasticity. {\em
  Mechanisms of Development} \textbf{126}, 942--949.

\bibitem{na2016grayscale}
Na JH, Bende NP, Bae J, Santangelo CD, Hayward RC. 2016  {Grayscale gel
  lithography for programmed buckling of non-Euclidean hydrogel plates}. {\em
  Soft Matter} \textbf{12}, 4985--4990.

\bibitem{Jones_2015}
Jones GW, Mahadevan L. 2015  Optimal control of plates using incompatible
  strains. {\em Nonlinearity} \textbf{28}, 3153--3174.

\bibitem{andrini2021theoretical}
Andrini D, Lucantonio A, Noselli G. 2021  A theoretical study on the transient
  morphing of linear poroelastic plates. {\em Journal of Applied Mechanics}
  \textbf{88}.

\bibitem{de2010pull}
De~Tommasi D, Puglisi G, Saccomandi G, Zurlo G. 2010  Pull-in and wrinkling
  instabilities of electroactive dielectric actuators. {\em Journal of Physics
  D: Applied Physics} \textbf{43}, 325501.

\end{thebibliography}

\end{document}